\documentclass[preprint,amsmath,amssymb,aps,pra,longbibliography,superscriptaddress]{revtex4-1}

\usepackage[margin=0.75in, top=1in, bottom=0.85in]{geometry}

\usepackage[final,letterspace=-0]{microtype}
\usepackage[colorlinks=true, linkcolor=blue, urlcolor=blue, citecolor=blue, anchorcolor=blue]{hyperref}

\usepackage{mathpazo} 
\usepackage{upgreek}
\usepackage{siunitx} 

\usepackage{graphicx}
\usepackage{dcolumn}
\usepackage{bm}
\usepackage{helvet}

\begin{document}

\title{Ultrafast space-time optical merons in momentum-energy space}

\author{Murat Yessenov}
\affiliation{CREOL, The College of Optics \& Photonics, University of Central Florida, Orlando, Florida 32816, USA}
\author{Ahmed H. Dorrah}
\affiliation{Harvard John A. Paulson School of Engineering and Applied Sciences, Harvard University, Cambridge, MA 02138, USA}
\author{Cheng Guo}
\affiliation{E. L. Ginzton Laboratory, Stanford University, Stanford, CA, USA}
\affiliation{Department of Electrical Engineering, Stanford University, Stanford, CA, USA}
\author{Layton A. Hall}
\affiliation{CREOL, The College of Optics \& Photonics, University of Central Florida, Orlando, Florida 32816, USA}
\author{Joon-Suh Park}
\affiliation{Harvard John A. Paulson School of Engineering and Applied Sciences, Harvard University, Cambridge, MA 02138, USA}
\author{Justin Free}
\affiliation{Micro-Photonics Laboratory, Center for Optical Materials Science and Engineering Technologies (COMSET), Holcombe Department of Electrical and Computer Engineering, Clemson University, 215 Riggs Hall, Clemson, SC 29634, USA}
\author{Eric G. Johnson}
\affiliation{Micro-Photonics Laboratory, Center for Optical Materials Science and Engineering Technologies (COMSET), Holcombe Department of Electrical and Computer Engineering, Clemson University, 215 Riggs Hall, Clemson, SC 29634, USA}
\author{Federico Capasso}
\affiliation{Harvard John A. Paulson School of Engineering and Applied Sciences, Harvard University, Cambridge, MA 02138, USA}
\author{Shanhui Fan}
\affiliation{E. L. Ginzton Laboratory, Stanford University, Stanford, CA, USA}
\affiliation{Department of Electrical Engineering, Stanford University, Stanford, CA, USA}
\author{Ayman F. Abouraddy}
\affiliation{CREOL, The College of Optics \& Photonics, University of Central Florida, Orlando, Florida 32816, USA}

\begin{abstract}
Skyrmions, topologically non-trivial localized spin structures, are fertile ground for exploring emergent phenomena in condensed matter physics and next-generation magnetic-memory technologies. Although magnetics and optics readily lend themselves to two-dimensional realizations of spin texture, only recently have breakthroughs brought forth three-dimensional (3D) magnetic skyrmions, whereas their optical counterparts have eluded observation to date because their realization requires precise control over the spatiotemporal spectrum. Here, we demonstrate the first 3D-localized optical skyrmionic structures with a non-trivial topological spin profile by imprinting meron spin texture on open and closed spectral surfaces in the momentum-energy space of an ultrafast optical wave packet. Precise control over the spatiotemporal spin texture of light -- a key requisite for synthesizing 3D optical merons -- is the product of synergy between novel methodologies in the modulation of light jointly in space and time, digital holography, and large-area birefringent metasurfaces. Our work advances the fields of spin optics and topological photonics, and may inspire new developments in imaging, metrology, optical communications, and quantum technologies.
\end{abstract}
\date{\today}
\maketitle

\newpage

Skyrmions are quasiparticles endowed with non-trivial yet stable topological spin textures that are localized in three dimensions (3D), and are immune to continuous deformations \cite{Skyrme1962NucPhys,TopologicalSolitonBook}. Arranging a continuous vector field on a sphere in a parametric space yields a wide variety of topological skyrmionic textures. In this form, skyrmions emerge in many physical systems, including nucleons \cite{Skyrme1962NucPhys}, Bose-Einstein condensates \cite{AlKhawaja2001, Han2017Book}, liquid crystals \cite{PhysRevLett.126.047801}, and magnetics \cite{Bogdanov2020NatPhysRev,Han2022}.  The latter particularly benefits from the localization and stability of skyrmions -- key ingredients for enabling future high-density, low-energy information storage and transfer technologies \cite{Fert2013NatNano}. Despite their rich topology in 3D space, the realizations of skyrmions to date have been mostly restricted to 'baby-skyrmions' -- whirls of spin configurations in a two-dimensional (2D) plane \cite{AlKhawaja2001, Nagaosa2013,Fert2017NatRevMat}. Realizing 3D skyrmions unlocks a vast domain of topological structures whose unique features may not only help increase the density of magnetic memory storage \cite{Gobel2021Review}, but may even lead to new phases of matter \cite{Cook2023JoP}. The search for physical systems to host such 3D topological configurations has been challenging \cite{Han2017Book}, and only recently have realizations emerged in fluid chiral ferromagnets \cite{Ackerman2017NatMat}, Bose-Einstein condensates \cite{Lee2018SciAdv}, and ferroelectric materials \cite{Das2019Nat}. In contrast to their well-studied 2D counterparts, the burgeoning study of 3D skyrmions is still witnessing rapid developments \cite{Tai2022NC}. 

Another recently explored platform for realizing skyrmions is that of optical waves, which is motivated by the potential for topologically structured electromagnetic fields to bring forth new applications, ranging from high-resolution imaging \cite{Du2019NP} to secure and resilient quantum communications \cite{Ornelas2024NP}. The first realizations of optical skyrmions were also restricted to 2D configurations, whereby topological structures are imprinted onto surface-bound stationary evanescent fields in plasmonic systems \cite{Tsesses2018Science,Dai2020Nat,Du2019NP}, and in freely propagating monochromatic fields at particular axial planes \cite{Gao2020PRA, Shen2022ACS,Marco2024ACSPhot}. Such 2D optical skyrmions are not 3D-localized, they are inherently static, and have not yet displayed topological stability over propagation. Similar restrictions apply to the latest realizations of 3D skyrmions in a monochromatic field in the form of hopfions \cite{Sugic2021NC,Wang2023PRL} whose structure is contained only within the focal volume of a lens but extends indefinitely in time. Fundamentally, the monochromaticity of previously reported optical skyrmions -- and thus their underlying 2D momentum space -- precludes their 3D localization and their transport.

Synthesizing fully localized optical skyrmions necessitates including the time dimension. Extending monochromatic 2D skyrmions directly into pulsed optical fields results in a trivial stretching into the time dimension (so-called skyrmion tubes \cite{Gobel2021Review}). 
Producing optical skyrmionic structures with non-trivial 3D-localized topology requires exploiting spatiotemporally non-separable pulsed optical fields \cite{Yessenov22AOP,shen2023JoO}, 
which has been proposed theoretically through sculpting the spin texture on a curved 2D spectral surface embedded in 3D momentum-energy space \cite{Guo2020PRL}. However, the challenge is that the spatiotemporal spectra for conventional pulsed optical beams inevitably occupy 3D \textit{volumes} -- rather than \textit{surfaces} -- in 3D momentum-energy space. This challenge has been recently addressed in the context of synthesizing propagation-invariant pulsed beams known as space-time wave packets (STWPs) \cite{Yessenov22AOP}. However, only simple spectral surfaces in the form of paraboloids of revolution have been realized to date \cite{Yessenov22NC,Yessenov22OL}. Unlocking the potential for topological skyrmionic structures in optics requires more sophisticated control over such spectral surfaces in momentum-energy space, ideally extending to never-before-realized \textit{closed} spectral surfaces on whose entirety a prescribed spin texture can be imprinted.

Here, we present the first 3D-localized optical skyrmionic spin texture imprinted onto open and closed 2D spectral surfaces embedded in the 3D momentum-energy space of an ultrafast pulsed optical beam. Following the proposal in \cite{Guo21Light}, we realize merons (half-skyrmions) that have been of particular interest in the context of momentum-energy space by virtue of their direct link to Berry curvature \cite{Loder2017PRB,Guo2020PRL} (Fig.~\ref{Fig:Concept}a). We call this class of skyrmionic structures henceforth space-time (ST) merons. An optical structure of this kind is localized in space and time and is endowed with topological stability over propagation at a controllable group velocity. These characteristics are produced by virtue of judicious coupling introduced between the spatial and temporal degrees of freedom of the optical field, which reduces the dimensionality of its spatiotemporal spectrum \cite{Yessenov22AOP}. The meron spin texture on these curved 2D spectral surfaces (Fig.~\ref{Fig:Concept}b) yields non-trivial 3D topological structures in space-time, whether using open (Fig.~\ref{Fig:Concept}c) or closed (Fig.~\ref{Fig:Concept}d) spatiotemporal spectral surfaces. To achieve these goals, we take two decisive steps that build upon previous experimental progress for synthesizing propagation-invariant STWPs \cite{Yessenov22AOP,Yessenov22NC}. First, we introduce a synthesis methodology to produce the first \textit{closed} spectral surfaces in optics via a novel two-to-one conformal coordinate transformation. Second, we have designed and fabricated a record-size birefringent dielectric metasurface consisting of $\sim10^{9}$ nano-fins, which serve as a symmetry-breaking device that introduces the target meron structure in momentum-energy space. ST merons synthesized via this methodology inherit some of the unique propagation characteristics of the underlying scalar STWP structure, such as transport at controllable group velocity \cite{Kondakci19NC}, anomalous refraction behavior at interfaces \cite{Bhaduri20NP}, and self-healing \cite{Kondakci18OL}. Thus, by modifying the geometry \textit{and} topology of spin distribution in momentum-energy space, the propagation dynamics of skyrmionic structures can be tuned in space-time, thereby opening a path for novel topological refraction phenomena \cite{Gao2018NP}, and new applications involving optical communications using higher dimension states of light \cite{Yang2020NP}, and topological light-matter interactions using photonic quasiparticles \cite{Rivera2020NatRevPhys}.


\section*{Theory of space-time merons in momentum-energy space}

\begin{figure*}[t!]
\centering
\includegraphics[width=17.5cm]{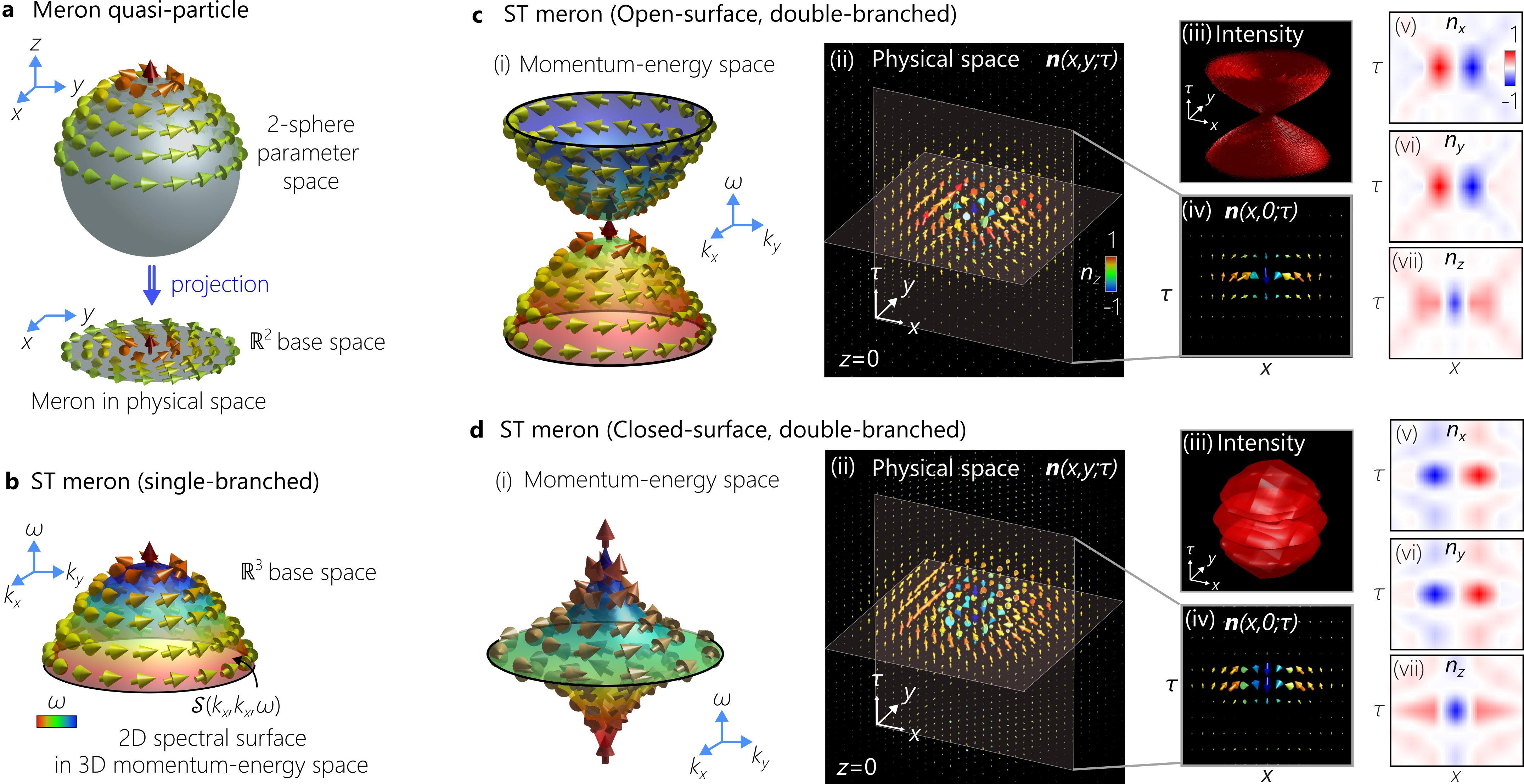}
\caption{\textbf{Ultrafast ST merons in momentum-energy space.} \textbf{a} The spin texture of a meron covers half the unit sphere in parameter space. For their implementations in physical space, the spin configuration is typically mapped from the parameter space onto the $\mathbb{R}^{2}$ base space via stereographic projection. \textbf{b} An ST meron is constructed by mapping the meron spin structure from the parameter space onto a 2D spectral surface $\mathcal{S}(k_{x},k_{y},\omega)$ embedded in the 3D momentum-energy $\mathbb{R}^3$ space of an optical wave packet. \textbf{c} The spin texture of an ST meron implemented on an open, dual-branch spectral surface in the form of a two-sheet paraboloid. \textbf{d} The spin texture of an ST meron implemented on a closed spectral surface in the form of a spinning-top. The plots in \textbf{c} and \textbf{d} show (i) the 3D spin distribution in momentum-energy space and (ii) in physical space $(x,y,z\!=\!0,\tau)$; (iii) the intensity isosurface $I(x,y,z\!=\!0;\tau)\!=\!0.1$; (iv) a 2D slice of the 3D spin distribution $\bm{n}(x,y=0,z=0;\tau)$ and (v-vii) its projections $n_x$, $n_y$ and $n_z$. The arrows in \textbf{a} indicate the spin orientation, and the arrows in \textbf{b-d} indicate the components of the Stokes vector. The colormaps on the spectral surface $\mathcal{S}(k_x,k_y,\omega)$ in \textbf{b}, \textbf{c}(i), and \textbf{d}(i) represent the distribution of frequencies $\omega$.}
\label{Fig:Concept}
\end{figure*}

\noindent
Skyrmions are represented by a continuous three-component unit vector field $\bm{n}(\bm{r})\!=\!(n_x(\bm{r}),n_y(\bm{r}),n_z(\bm{r}))$ on a unit 2-sphere parameter space; where $\bm{r}\!=(x,y,z)$ and $x^2\!+\!y^2\!+\!z^2\!=\!1$. These configurations are distinguished by a non-zero topological Skyrmion number $N$~\cite{Nagaosa2013}, which quantifies the number of times $\bm{n}(\bm{r})$ wraps around the unit sphere. Merons, or half-skyrmions, carry a skyrmion number $N=\pm \frac{1}{2}$, indicating that  $\bm{n}(\bm{r})$ covers only half the unit sphere, from the north pole to the equator (Fig.~\ref{Fig:Concept}a). The meron texture is explicitly given in polar coordinates by: $\bm{n}(\rho,\phi)\!=\!\frac{1}{\sqrt{\zeta^{2}+\Delta^{2}}}\left\{\rho\cos (\phi+\gamma),\rho\sin(\phi+\gamma),\Delta\right\}$; where $\rho\!=\!\sqrt{{x}^{2}+y^{2}},\quad\phi= \arg(x+iy)$, the parameter $\Delta$ specifies the meron radius, and the helicity $\gamma\in(-\pi, \pi]$ determines the chirality of the spin texture, which gives rise to a classification of merons into N{\'e}el-type ($\gamma\!=\!0$) and  Bloch-type ($\gamma\!=\!\pm\pi/2$) \cite{Nagaosa2013}. 
Previous experimental realizations of optical skyrmions have typically relied on modifying the spatial profile of a monochromatic optical field \cite{Beckley2010OE,Tsesses2018Science,Du2019NP,Dai2020Nat,Shen2022ACS, Sugic2021NC, Shen2023NP}, by mapping the 3D spin vector field $\bm{n}(\bm{r})$ from the parameter space to a flat 2D surface disk in $\mathbb{R}^{2}$ as a base space (real space), via a Hopf mapping (hopfions) or a stereographic projection (2D skyrmions) \cite{Nagaosa2013}; see Fig.~\ref{Fig:Concept}A.

In contrast, here we make use of a curved 2D spectral surface $\mathcal{S}(k_{x},k_{y},\omega)$ as a base space in the $\mathbb{R}^3$ momentum-energy domain of a pulsed optical field in narrowband paraxial regime (Fig.~\ref{Fig:Concept}b,c,d); where $k_{x}$ and $k_{y}$ are the transverse components of the wave vector, and $\omega$ is the temporal frequency. We map the meron texture from the hemisphere in parameter space to the spin (polarization) on $\mathcal{S}(k_{x},k_{y},\omega)$, such that $\bm{n}(\bm{r})\!=\!\left(s_{1}(\bm{r}),s_{2}(\bm{r}),s_{3}(\bm{r})\right)$; here $s_{1}$, $s_{2}$, and $s_{3}$ are normalized Stokes parameters. Scalar optical fields restricted to simple 2D spectral surfaces are known as STWPs \cite{Kondakci17NP,Yessenov22AOP}, whose characteristics are useful for a variety of applications \cite{Yessenov22AOP,shen2023JoO}. First, note that a one-to-one relation enforced between the transverse radial wave number $k_{r}\!=\!\sqrt{k_{x}^2\!+\!k_{y}^2}$ and the temporal frequency $\omega$ leads to diffraction-free propagation \cite{Yessenov22AOP}, thus potentially enabling the transportation of localized topological structures over extended distances \cite{Kondakci17NP}. Moreover, when the space-time coupling takes the form of a paraboloid $\tfrac{\Omega(k_r)}{\omega_{\mathrm{o}}}\!\approx\!\tfrac{k_{r}^{2}}{2k_{\mathrm{o}}^{2}(1-c/\widetilde{v})}$ (Fig.~\ref{Fig:Concept}b), the optical wave packet travels rigidly at a controllable group-velocity $\widetilde{v}$ in linear optical media \cite{Guo21Light,Yessenov22NC}; here $\omega_{\mathrm{o}}$ is a carrier frequency, $k_{\mathrm{o}}\!=\!\tfrac{\omega_{\mathrm{o}}}{c}$ is the associated wave number, $c$ is the speed of light in vacuum, and $\Omega\!=\!\omega-\omega_{\mathrm{o}}$ (Fig.~\ref{Fig:Concept}b). The spatiotemporal spectrum of an azimuthally symmetric STWP can be represented by $\widetilde{E}(k_{r},\Omega)\!=\!\widetilde{E}(k_{r})\delta(\Omega-\Omega(k_r))$, where $\widetilde{E}(k_{r})$ is the spatial spectrum, and the delta function is relaxed in practice and replaced with a narrowband spectral function \cite{Yessenov22NC}. After incorporating the meron spin texture, the spatiotemporal spectrum of such ST merons in the circular polarization basis is: $\widetilde{\bm{E}}(k_{r},\phi_{k},\Omega)\!=\widetilde{E}(k_{r},\Omega)[\sin \frac{\theta(k_{r})}{2},-e^{i\chi(\phi_k)} \cos \frac{\theta(k_{r})}{2}]^{\mathrm{T}}$; here, `${\mathrm{T}}$' is the transpose operation, $\theta\left(k_{r}\right)\!=\!\pi-\arccos \frac{\kappa}{\sqrt{k_{r}^{2}+\kappa^{2}}}$, $\chi \left(\phi_k \right)\!=\!\phi_k + \gamma - \pi$, and $\phi_{k}\!=\!\arg(k_{x}+ik_{y})$ (see Supplementary Section~I), and we set $\gamma\!=\!\pi/2$ to produce a Bloch-type topological structure. The vectorial field profile in physical space takes the form $\bm{E}(r,z;\tau)\!\propto\!\iint\!dk_{r}d\Omega\;\;\;k_{r}\widetilde{\bm{E}}(k_{r},\Omega)J_{0}(k_{r}r)e^{-i\Omega\tau}\!=\!\bm{E}(r,0;\tau)$, where $J_{0}(\cdot)$ is the $0^{\mathrm{th}}$-order Bessel function of the first kind \cite{Yessenov22NC}, and $\tau\!=\!t-z/\widetilde{v}$ is a time frame moving at the group velocity $\widetilde{v}$.

Using the single-branched paraboloidal spectral surface as a building block, we construct complex surface geometries by modifying the space-time coupling function $\Omega(k_{r})$. We consider two such spectral structures: an open-surface double-branched paraboloid $\Omega(k_{r})\!=\!\pm(ak_{r}^{2}+b)$, (Fig.~\ref{Fig:Concept}c), and a closed-surface spinning-top $\Omega(k_r)\!=\!\pm(\frac{a}{k_{r}}+b)$ (Fig.~\ref{Fig:Concept}d); where $a$ and $b$ are positive real constants. The meron texture here is the same on both branches of these surfaces in momentum-energy space (Fig.~\ref{Fig:Concept}c,d, panel (i)). In principle, however, independent spin profiles could be prescribed on the two branches to explore other skyrmionic structures. The structure of the ST merons in physical space $\bm{E}(x,y,z;\tau)$ is calculated via a 3D Fourier transform (at $z\!=\!0)$. First, the 3D structure of the spin vector $\bm{n}(x,y,z;\tau)$ for both ST merons reveals non-trivial spin distributions localized in the 3D space-time volume of the wave packet; see Fig.~\ref{Fig:Concept}c,d, panel (ii). This is in striking contrast with the spectral distributions in momentum-energy space, where the spin vector is distributed over only the surface $\mathcal{S}(k_x,k_y,\omega)$. The 3D plots of the intensity isosurface $I\!=\!0.1$ (after normalizing the intensity to unity) in panel (iii), reveal that the ST meron corresponding to the closed spectral surface is more localized (Fig.~\ref{Fig:Concept}d), whereas the open-surface ST meron shows the characteristic X-shaped structure of free-space STWPs (Fig.~\ref{Fig:Concept}c) \cite{Kondakci17NP,Yessenov22AOP}. To clearly visualize the spatiotemporal structure of the spin texture in physical space, we plot the spin vector $\bm{n}(x,y=0,z=0;\tau)$ in panel~(iv), and the spin projections $n_x$, $n_y$, and $n_z$ in panels~(v-vii) in the 2D space-time plane $(x,\tau)$. These plots reveal a swerve of the spin in the ST meron on a spatial scale smaller than the beam size, and on a shorter temporal scale that the pulse width, corresponding to the sub-wavelength spatial features of 2D skyrmions in evanescent fields \cite{Du2019NP}.

\section*{Experimental setup}

\begin{figure}[t!]
\centering
\includegraphics[width=16cm]{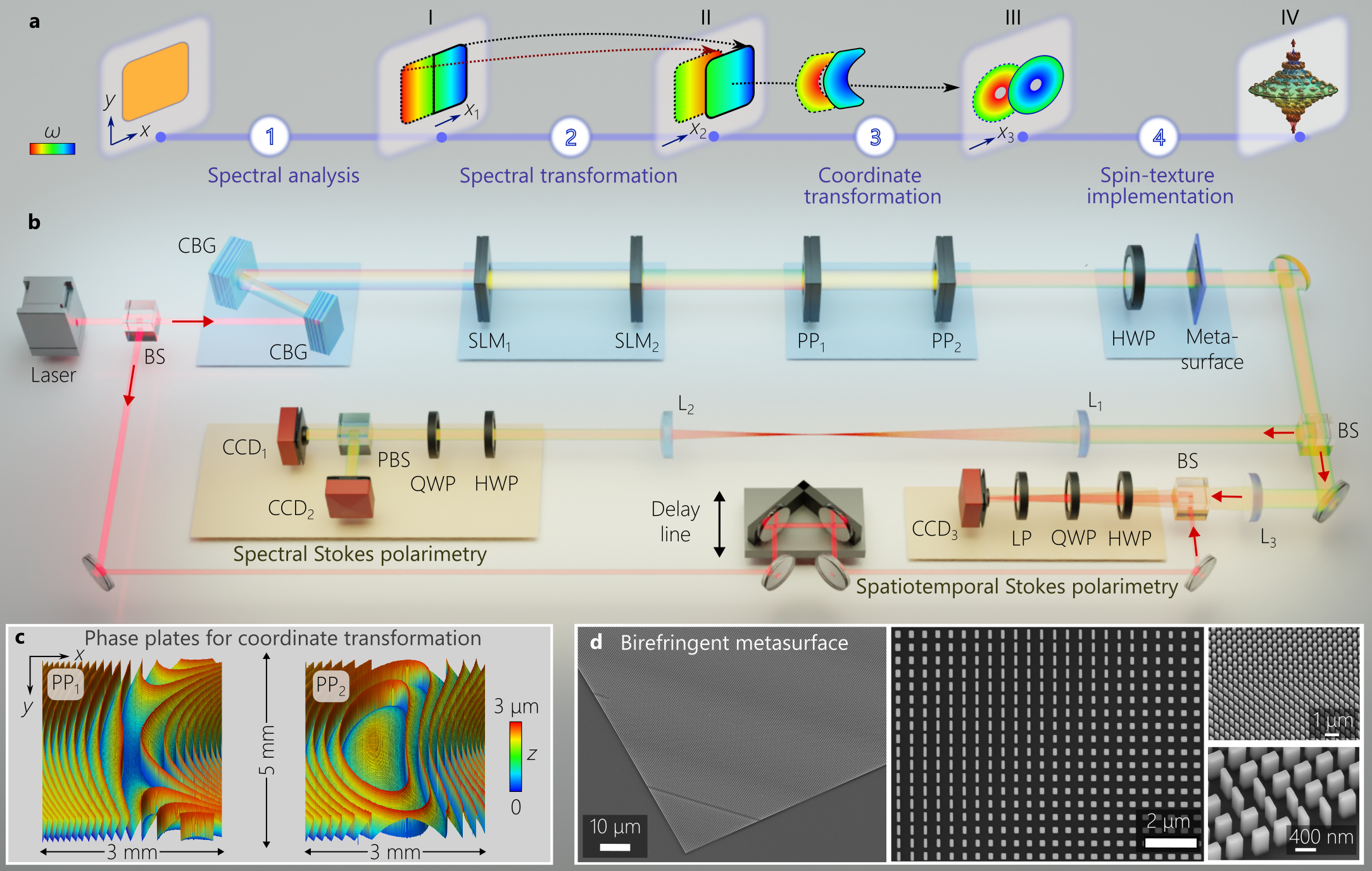}
\caption{\textbf{Schematic of the experimental setup to synthesize and characterize ST merons.} \textbf{a} Evolution of the field structure through four stages of the ST meron synthesis system. \textbf{b} Experimental setup for synthesizing and characterizing ST merons. In the first stage, the spectral components of an incoming ultrafast pulse are spatially separated via a pair of chirped volume Bragg gratings (CBGs) to introduce a linear spatial chirp $x_{1}({\Omega})$ (inset I in \textbf{a}). In the second stage, a two-to-one spectral transformation $|x_{1}|\!\rightarrow\!x_{2}$ implemented via a pair of spatial light modulators (SLMs) maps a pair of spectral lines at the transverse positions $\pm x_{1}$ to the same transverse position $x_{2}$, producing two spatially overlapping spectra $x_{2}(|\Omega|)$ (inset II in \textbf{b}). A conformal coordinate transformation stage (implemented by a pair of phase plates PP$_{1}$ and PP$_{2}$) transforms the linear chirp into a radial chirp $r(|\Omega|)$ (inset III in \textbf{a}). A birefringent metasurface then imparts the target polarization texture $\bm{n}(r(\Omega))$ to the radially spread spectrum (inset IV in \textbf{a}). The spatiotemporal structure of the ST meron is characterized in the spectral domain (after the $4f$-system comprising lenses $\mathrm{L}_1$ and $\mathrm{L}_2$) via spectral Stokes polarimetry, and in real space (after the Fourier-transforming lens $\mathrm{L}_3$) by performing spatiotemporally resolved Stoked polarimetry. \textbf{c} Surface profilometry micrographs of the central part of the phase plates (PP$_{1}$ and PP$_{2}$). \textbf{d} Scanning electron microscope (SEM) micrographs of the birefringent dielectric metasurface at different spatial scales.} 
\label{Fig:Setup}
\end{figure}

\noindent
The key for successfully synthesizing ST merons in momentum-energy space is twofold: sculpting a spectral surface $\mathcal{S}(k_{x},k_{y},\Omega)$ through introducing a space-time coupling $\Omega(k_r)$ into a pulsed optical field, \textit{and} exercising high-efficiency, point-by-point control over the spin profile on this spectral surface. This task is particularly challenging in the case of a closed spectral surface. Our strategy for synthesizing ST merons from a generic optical pulse follows the four-stage scheme illustrated in Fig.~\ref{Fig:Setup}a comprising: (1) spectral analysis; (2) spectral transformation; (3) coordinate transformation; and (4) spin-texture implementation. To construct a prescribed spectral surface, the first step is to spatially resolve the incoming spectrum (spectral analysis). At this point, the field is endowed with linear spatial chirp $x_{1}(\Omega)$, whereby the frequencies are arranged in a fixed sequence along the $x_{1}$ axis (inset I). Realization of a prescribed spectral surface (such as that in Fig.~\ref{Fig:Concept}b) requires arranging the frequencies in a particular sequence $x_{1}(\Omega)\!\rightarrow\!x_{2}(\Omega)$ (spectral transformation). Moreover, achieving double-branched open or closed spectral surfaces (Fig.~\ref{Fig:Concept}c,d) necessitates a two-to-one spectral mapping $\pm x_{1}\!\rightarrow\!x_{2}$, thereby re-positioning two spectral lines at two distinct locations at the input to one location at the output, which is in contrast to previous realizations of optical conformal mappings that have \textit{all} been one-to-one. The result is two precisely overlapped spatially resolved spectra with opposite chirp rates $x_{2}(|\Omega|)$ that were initially contiguous (inset II). The last step for constructing the spectral surface $\mathcal{S}(k_{x},k_{y},\Omega)$ is the coordinate transformation that maps the rectilinear chirp $x_2(|\Omega|)$ into a radial chirp $r(|\Omega|)$; i.e., lines corresponding to different wavelengths at the input are converted into circles at the output plane (inset III). Finally, imprinting the meron spin texture onto the radially resolved spectrum generates the ST meron in momentum-energy space (inset IV).



\cite{Dai2020Nat}

The setup for synthesizing ST merons from a generic optical pulse is depicted in Fig.~\ref{Fig:Setup}b. Starting with linearly polarized pulses from a mode-locked Ti:Sapphire laser of width 100~fs, central wavelength $\lambda_{\mathrm{o}}\approx800$~nm, bandwidth $\Delta\lambda\approx10$~nm, and beam width $\approx3$~mm, the spectrum is spatially resolved (spectral analysis) via a pair of chirped volume Bragg gratings (CBGs) \cite{Kaim13SPIE}. The first CBG spatially spreads the spectrum (spatial chirp) of the obliquely incident pulse, but simultaneously stretches the pulse in time (temporal chirp). A second identical but reversed-orientation CBG eliminates the temporal chirp while doubling the spatial chirp (see Methods) \cite{Yessenov22NC}. 

The second stage comprises a pair of spatial light modulators (SLMs) that impart two phase distributions to implement an afocal conformal spatial transformation from the input plane $x_1$ to the output plane $x_{2}\!=\!f(x_{1})$ \cite{Bryngdahl74JOSA,Hossack87JOMO}. Because the spectrum is spatially resolved at its input plane $x_{1}(\Omega)$, such a spatial transformation corresponds to a spectral transformation. This mapping modifies three aspects of the resolved spectrum. First, it pre-compensates for an unwanted scaling occurring in a subsequent stage (coordinate transformation). Second, it enables the realization of any prescribed spectral structure in real time by tuning the transformation function $f(x_{1})$. Third, it allows us to implement a two-to-one spectral mapping $x_{2}(|\Omega|)$, thus producing the surfaces in Fig.~\ref{Fig:Concept}c,d. To perform this mapping, two phase distributions are concatenated on the first SLM to map the two halves of the input field in the input plane onto the same spatial domain in the output plane $f(|x_{1}|)$, and the two phase patterns at the output plane are interleaved on the second SLM (Supplementary Section II.B). Consequently, each position in the output plane is now associated with two wavelengths, i.e., $x_{2}(\Omega)\!=\!x_{2}(-\Omega)$, whereas each position in the input plane was associated with one wavelength $x_{1}(\Omega)$. This is a key attribute needed for constructing two-branched spectral surfaces in momentum-energy space. 

In the third stage (coordinate transformation), a pair of phase plates implements another conformal mapping in the form of a log-polar coordinate transformation that maps each vertical line at its input plane into a circle at its output \cite{Bryngdahl74JOSA, Berkhout10PRL}. 
Restricting the implementation of such a coordinate transformation to two phase plates (rather than three or more) necessitates exponentiating the radial coordinate at the output plane \cite{Bryngdahl74JOSA}, which is pre-compensated by introducing the logarithmic scaling in the preceding spectral transformation stage, as discussed above. The conjunction of spectral and spatial transformations produces two spatially overlapping spectra that are spread radially with oppositely signed radial chirps $r(|\Omega|)$; see Supplementary Section II.C. Success in producing closed spectral surfaces in momentum-energy space is hindered by an intrinsic feature of the log-polar coordinate transformation; namely, a singularity at the center of the second phase distribution that leads to a null at the center of the spatiotemporal spectrum at $k_{r}\!=\!0$. Minimizing the singularity region while remaining in the paraxial regime necessitates fabricating large-area phase plates ($14\times14$~mm$^{2}$ here), which are produced using an analog photolithographic process \cite{Sung2005OL} (Fig.~\ref{Fig:Setup}c; Supplementary Section II.C).

Lastly, realizing the meron spin texture requires a wavefront shaping platform that imparts a point-by-point birefringent response. Dielectric metasurfaces are an ideal candidate for this task by virtue of their sub-wavelength resolution, compact footprint, and highly efficient polarization-dependent modulation of the phase and amplitude at each point in an optical field by utilizing form birefringence, while operating away from resonances to mitigate errors arising from fabrication inaccuracies \cite{Rubin21AOP, Dorrah22ScienceRev}. We introduce the metasurface at this point in the synthesis procedure to impart to the incident linearly polarized, spatially resolved double-spectrum a spatially varying polarization profile corresponding to the spin texture of a meron. The large area of the phase plates used in the coordinate transformation stage necessitates in turn fabricating a record-size metasurface comprising $\approx10^{9}$ unit cells of $520\times520$~nm$^{2}$ area each, containing an 800-nm-tall nanofin of TiO$_{2}$ on a SiO$_{2}$ substrate. The geometry of each meta-atom in the metasurface is tuned to achieve form birefringence (Fig.~\ref{Fig:Setup}d). We adopt a phase-retrieval-like algorithm \cite{Yu11Science,Kamali18NanoP,Rubin21AOP,Dorrah22ScienceRev} to realize the requisite far-field complex amplitude profiles for two orthogonal polarization components \cite{Arrizon07JOSAA}, which yields the requisite parameters of each unit cell. The metasurface was produced via standard planar fabrication processes (see Methods), and the measured spatially resolved Stokes parameters confirm that it imparts the target spin texture (Supplementary Fig. S11). Finally, a Fourier-transforming lens converts this double-valued spatially resolved spectrum endowed with spin texture from momentum-energy space $(k_{x},k_{y},\omega)$ to physical space spanned by $(x,y,z;\tau)$.


\section*{Measurement results}

\begin{figure}[t!]
\centering
\includegraphics[width=14cm]{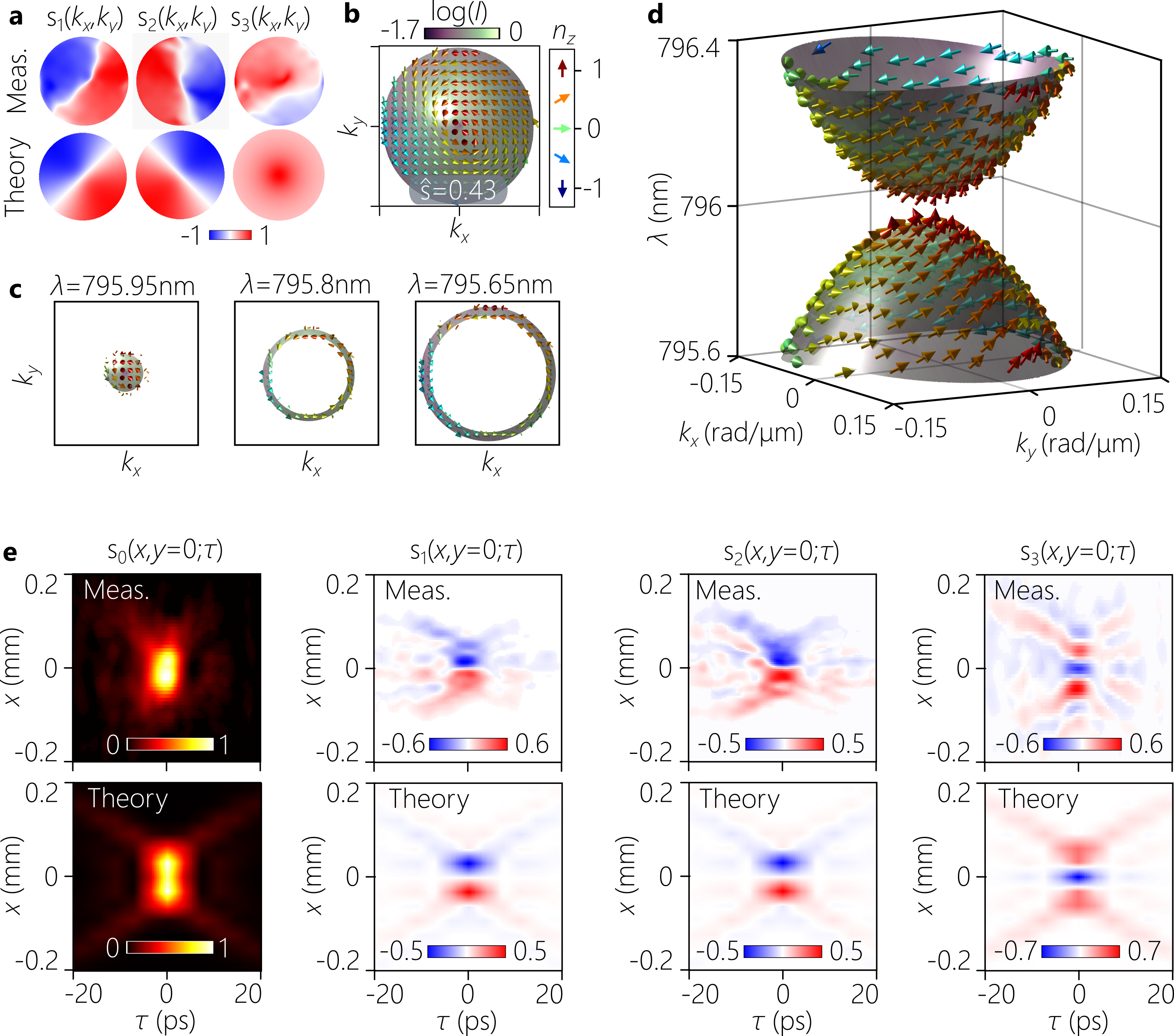}
\caption{\textbf{Characterizing an ST meron whose spin texture is imprinted on a two-sheet paraboloid spectral surface}. \textbf{a} Measured (top row) and calculated (bottom row) spectral Stokes parameters $s(k_{x},k_{y})$ in  $(k_{x},k_{y})$-space (integrated over all temporal frequencies $\omega)$. \textbf{b} A plot of the spin texture (arrows) reconstructed from the measurements in \textbf{a}, overlaid with the measured spectral intensity profile $I(k_{x},k_{y})$ measurements (colormap). \textbf{c} Cross-sectional plots of \textbf{b} at different iso-frequency planes corresponding to $\lambda\!=\!795.95$~nm, 795.8~nm, and 795.65~nm. \textbf{d} Measured spin texture $\bm{n}(k_x,k_y,\lambda)$ (arrows) plotted in the 3D spatiotemporal spectral domain (momentum-energy space), plotted over the ST spectral surface overlaid with the measured spectral intensity profile $I(k_{x},k_{y})$. \textbf{e} Spatiotemporal Stokes parameters $(s_{0},s_{1},s_{2},s_{3})$ at a fixed axial plane $z\!=\!0$, where $s_{0}$ corresponds to the total intensity. The panels are plotted in the $(x,\tau)$ plane corresponding to $y\!=\!0$. The first row presents the measurements and the second row the corresponding theoretical predictions. In \textbf{d} the measured temporal bandwidth is $\Delta\lambda\!\approx\!0.8$~nm, and the spatial bandwidth is $\Delta k_{r}\!\approx\!0.18$~rad/mm; in \textbf{e} the measured pulse width at the beam center is $\Delta\tau\!\approx\!6$~ps, and the beam size at the pulse center is $\Delta x\!\approx\!56$~$\mu$m.}
\label{Fig:DataParabola}
\end{figure}

\begin{figure}[t!]
\centering
\includegraphics[width=14cm]{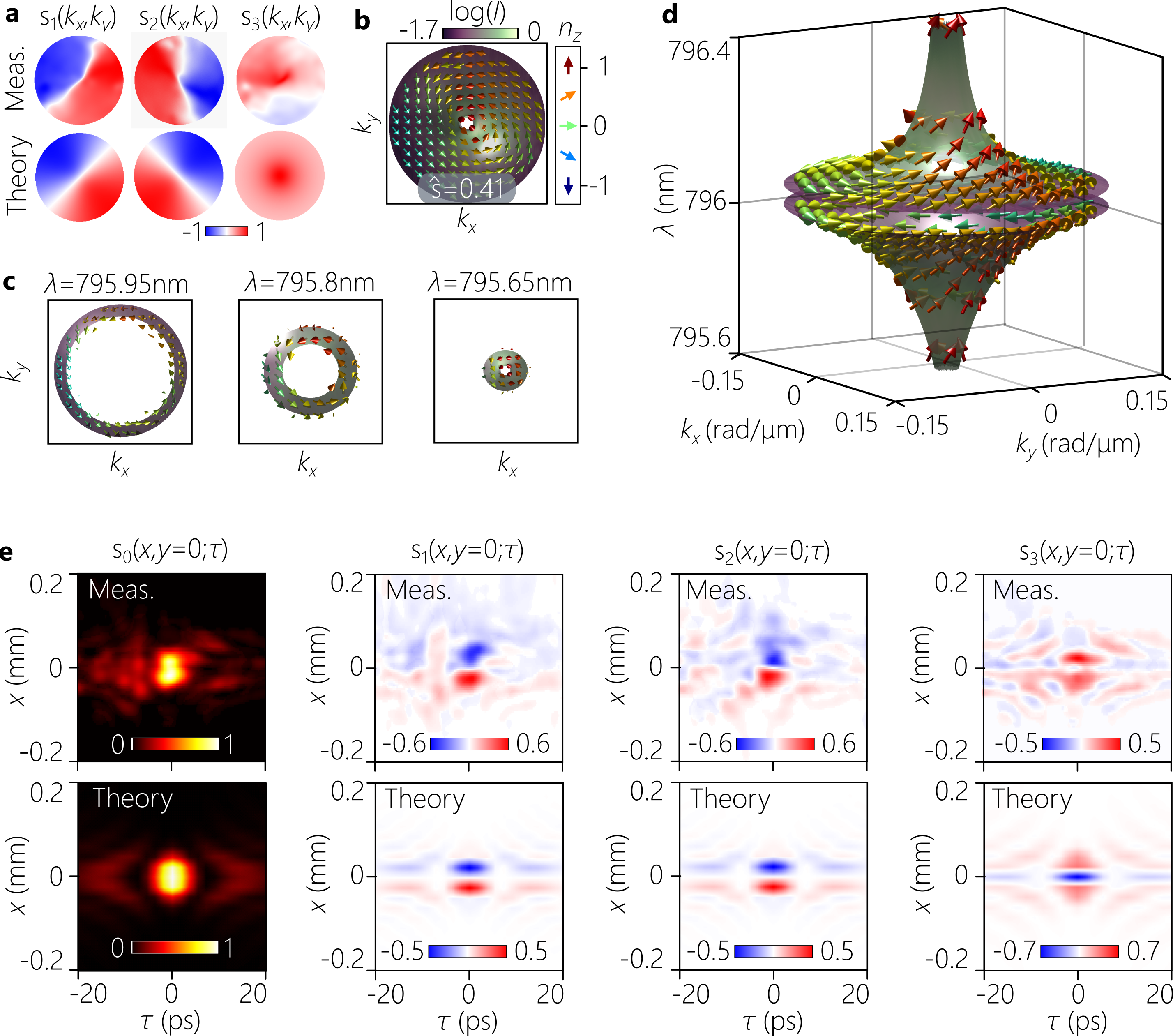}
\caption{\textbf{Characterization of an ST meron whose spin texture imprinted on a spinning-top spectral surface.} 
\textbf{a} Measured (top row) and calculated (bottom row) spectral Stokes parameters $s(k_{x},k_{y})$ in  $(k_{x},k_{y})$-space. \textbf{b} A plot of the spin texture (arrows) reconstructed from the measurements in \textbf{a}, overlaid with the measured spectral intensity profile $I(k_{x},k_{y})$ measurements (colormap). \textbf{c} Cross-sectional plots of \textbf{b} at different iso-frequency planes corresponding to $\lambda\!=\!795.95$~nm, 795.8~nm, and 795.65~nm. \textbf{d} Measured spin texture $\bm{n}(k_x,k_y,\lambda)$ (arrows) plotted in the 3D spatiotemporal spectral domain (momentum-energy space), plotted over the ST spectral surface overlaid with the measured spectral intensity profile $I(k_{x},k_{y})$. \textbf{e} Spatiotemporal Stokes parameters $(s_{0},s_{1},s_{2},s_{3})$ at a fixed axial plane $z\!=\!0$, where $s_{0}$ corresponds to the total intensity. The panels are plotted in the $(x,\tau)$ plane corresponding to $y\!=\!0$. The first row presents the measurements and the second row the corresponding theoretical predictions. In \textbf{d} the measured temporal bandwidth is $\Delta\lambda\!\approx\!0.8$~nm, and the spatial bandwidth is $\Delta k_{r}\!\approx\!0.17$~rad/mm; in \textbf{e} the measured pulse width at the beam center is $\Delta\tau\!\approx\!6.6$~ps, and the beam size at the pulse center is $\Delta x\!\approx\!63$~$\mu$m.
}
\label{Fig:DataSpinningTop}
\end{figure}

\noindent
We carried out the measurements for the two ST merons in Fig.~\ref{Fig:Concept}c,d: 
the \textit{open}-surface ST meron with a spatiotemporal spectrum comprising oppositely oriented paraboloids of revolution (Fig.~\ref{Fig:DataParabola}), and the \textit{closed}-surface ST meron with the spectrum that takes the shape of a spinning top (Fig.~\ref{Fig:DataSpinningTop}). Switching between these two spectral surfaces and modifying their parameters requires only changing the SLM phase distributions in the spectral transformation stage.

We first characterize the spin texture of ST merons in momentum space $(k_x,k_y)$ by spatially resolved Stokes polarimetry carried out before the Fourier-transforming lens $\mathrm{L}_{3}$ in Fig.~\ref{Fig:Setup}b, which reveals the momentum-space Stokes parameters $s_{i}(k_{x},k_{y})$, $i=1,2,3$ (Fig.~\ref{Fig:DataParabola}a and Fig.~\ref{Fig:DataSpinningTop}a). From these measurements, we construct the spin distribution $\bm{n}(k_{x},k_{y})$ for two ST merons (arrows in Fig.~\ref{Fig:DataParabola}b and Fig.~\ref{Fig:DataSpinningTop}b) and plot them overlaying the measured spectral intensity $I(k_{x},k_{y})$ (colormap in Fig.~\ref{Fig:DataParabola}b and Fig.~\ref{Fig:DataSpinningTop}b). From these measurements, we extract Skyrmion numbers of $N\!=\!0.41$ and $N\!=\!0.42$ for the open- and closed-surface ST merons, respectively, which are close to the target value of $N\!=\!0.416$ (see Supplementary Section~III.C). An additional measurement of the correlation between the radial wave number $k_{r}\!=\!\sqrt{k_{x}^{2}+k_{y}^{2}}$ and the temporal frequency $\omega$ allows us to isolate the wavelength dependence of these Stokes parameters (Fig.~\ref{Fig:DataParabola}c and Fig.~\ref{Fig:DataSpinningTop}c), thereby enabling the reconstruction of the ST meron spin structure over the entire spectral surface (Fig.~\ref{Fig:DataParabola}d and Fig.~\ref{Fig:DataSpinningTop}d); see Supplementary Section~III. 

In addition, we have characterized the spin texture of the ST merons in physical space by measuring the spatiotemporally resolved Stokes parameters after the Fourier-transforming lens $\mathrm{L}_{3}$ via linear interferometry (Fig.~\ref{Fig:Setup}b). The ST meron synthesis setup is placed in one arm of a Mach-Zehnder interferometer, and the initial 100-fs pulses are employed as a reference that traverses an optical delay line $\tau$ placed in the other interferometer arm. The visibility of the spatially resolved interference fringes recorded by a camera placed in the common path helps reconstruct the spatiotemporal intensity profile of the ST meron at a fixed relative delay $\tau$ \cite{Kondakci19NC}. Sweeping the delay line $\tau$ thus yields the full spatiotemporally resolved intensity profile $I(x,y,z;\tau)$ at the observation plane $z\!=\!0$ (Fig.~\ref{Fig:DataParabola}e and Fig.~\ref{Fig:DataSpinningTop}e, $s_{0}(x;\tau)$). The visibility of the interference fringes is maximized only after projecting the combined light fields onto one polarization state. Therefore, by repeating the procedure for different polarization components we retrieve the spatiotemporally resolved Stokes parameters  $s_{j}(x,y,z\!=\!0;\tau)$ for the ST merons, $j\!=\!1,2,3$.  (Fig.\ref{Fig:DataParabola}e and Fig.~\ref{Fig:DataSpinningTop}e). We plot the 2D intensity profiles and the Stokes parameters in the $(x,\tau)$-plane corresponding to $y\!=\!0$ for clarity, as we have done in Fig~\ref{Fig:Concept}c,d, without loss of crucial information. These measurements are in good agreement with the theoretical predictions plotted in Fig.~\ref{Fig:DataParabola}e and Fig.~\ref{Fig:DataSpinningTop}e (second row), corresponding to $s_{1}$, $s_{2}$ and $s_{3}$. The measured spatiotemporal evolution of spin texture is observed within a localized region of $\approx6$~ps temporal width and $\approx60$~$\mu$m transverse spatial width. It is clear that ST merons with an identical topological spin structure in momentum-energy space (produced by the metasurface) can be endowed with distinct spatiotemporal spin configuration in physical space by virtue of the modified spectral surface (Fig.~\ref{Fig:DataParabola}d and Fig.~\ref{Fig:DataSpinningTop}d). This suggests that ST merons may enable on-demand modulation of transverse spin profiles at ultrafast speeds within a localized transverse space, which may be useful for optical sensing and data storage applications \cite{Shen2023NP}. Note that the spatiotemporal profile of the open-surface ST meron is X-shaped as is characteristic of STWPs in free space (Fig.~\ref{Fig:DataParabola}e), and that of the closed-surface ST meron is circularly symmetric with respect to space and time, which is characteristic of closed spatiotemporal spectral supports (Fig.~\ref{Fig:DataSpinningTop}e), as recently verified for the restricted-dimensional case of light sheets \cite{Hall23NatPhys}.

\section*{Discussion}
\noindent
We have synthesized optical ST merons in which the topological spin texture is imprinted on open and closed 2D spatiotemporal spectral surfaces embedded in the 3D momentum-energy space of a pulsed optical beam, and have observed the associated 3D spin texture in space-time. Our work paves the way for a new class of ultrafast photonic quasiparticles with rich nanoscale topology that can be exploited in sensing, microscopy, secure information transfer, and light-matter interactions. This work also raises new questions and suggests avenues for further development. First, by providing independent control over the spin texture associated with the two branches of the spectral surface (which spatially overlap after the spectral transformation stage), other topological quasi-particles besides merons can be readily produced. Introducing the spin texture at the spectral transformation stage (Fig.~\ref{Fig:Setup}b), via polarization-insensitive SLMs or phase plates in lieu of polarization-sensitive SLMs, provides access to the entire gamut of topological spin textures. Second, further pushing the limits of optical conformal mappings in momentum-energy space may yield more complex spectral surfaces on which the spin texture is implemented. Of special interest are spectral surfaces that are themselves endowed with non-trivial topological structure, such as tori or nested tori, which can thus give rise to space-time hopfions \cite{Kent2021NC,Wang2023PRL} and Shankar's skyrmions \cite{Lee2018SciAdv}. Third, our methodology can be directly extended to add orbital angular momentum (OAM) to each wavelength, sub-band, or the entire spectrum of the ST meron, which enables the exploration of topological spin-orbit coupling \cite{Bliokh2015NP,Bliokh2015Science} and the investigation of recently discovered effects, such as optical self-torque \cite{Rego2019Science} and time-varying OAM \cite{Chen2022AdvSc, Su23ArXiV}, in the context of optical skyrmions. Fourth, the versatility of optically synthesizing 3D spin texture may be the basis for transferring this spin texture from light to matter \cite{Dai2022NatRevPhys, Rivera2020NatRevPhys}, thereby exciting topological quasi-particle structures in material systems; for example, by laser-induced magnetization \cite{Beruto2018PRL} for next generation memory devices and spintronics \cite{Basini2024Nat}. This requires examining the synthesis of the spin texture directly in 3D physical space rather than momentum-energy space. Fifth, the optical ST merons demonstrated here may potentially be used to excite propagating surface plasmon-polariton waves endowed with skyrmionic texture \cite{Ichiji2023PRA}, which enables field confinement in the deep sub-wavelength regime, and may enable ultrafast spontaneous emission and few-molecule strong coupling, or give rise to new phenomena such as multiphoton spontaneous emission and forbidden transitions \cite{Rivera2020NatRevPhys}. Moreover, precise control over the 3D polarization configuration of an optical wave packet has direct influence on nonlinear optical effect \cite{Wu:22}.

Additionally, the ST merons developed here raise the prospect of potential coupling to optical fibers and waveguides \cite{Shiri20NC}. In this respect, closed spectral surfaces in momentum-energy space offer an advantage because they always result in pulsed beams that experience effective normal group-velocity dispersion in free space \cite{Hall23NatPhys}. Consequently, such wave packets are dispersion-free in optical fibers in the telecommunications window where the group-velocity dispersion is anomalous. Finally, considering recent demonstrations of using space-time coupled fields for speckle-resistance \cite{Diouf22SA} and self-healing after an obstruction \cite{Kondakci18OL}, it is intriguing to determine the degree of topological resilience of optical ST merons against external perturbations in complex media, which has not been conclusively studied yet. Such topological protection against scattering, surface defects, and perturbations is crucial for using optical 3D ST topological quasi-particles as carriers of information in optical communications and beyond. 

\clearpage

\section*{Methods}
\subsection*{Spectral analysis via a chirped Bragg grating (CBG)}
We spatially resolve the spectrum by making use of a chirped volume Bragg grating (CBG) whose grating period varies longitudinally. When an optical pulse is incident onto the CBG, each wavelength reflects from a different depth within it where the Bragg condition is met. Therefore, when the field is incident normally onto the CBG, a group delay is introduced between the different temporal frequencies (temporal chirp), thus yielding a stretched pulse. This property of CBGs is widely used for pulse stretching and compression in high-power chirped pulse amplification systems in light of the high damage threshold, high efficiency, and potential for introducing extremely large temporal chirps. A less-known property of CBGs is their ability to introduce a spatial chirp -- resolving the spectrum spatially -- along with a temporal chirp when the pulse is incident obliquely on a conventional CBG, or incident normally on a so-called rotated CBG  (r-CBG) in which the Bragg structure is rotated with respect to the input facet. For our purpose, it is crucial to remove the temporal chirp while retaining the spatial chirp. This is achieved by directing the field emerging from the CBG to an identical CBG placed in a reversed geometry with respect to the first; see Supplementary Section II.A and Supplementary Fig. S3. In our experiment we made use of a CBG (OptiGrate L1-021) with central Bragg period $\Lambda_{\mathrm{o}}\!=\!270$~nm, spectral chirp rate $\beta\!=\!-30$~pm/mm, average refractive index of $n\!\approx\!1.5$, and length $L\!=\!34$~mm. The input beam is incident at an angle $\phi\!\approx\!16^{\circ}$ with respect to the normal to the CBG entrance surface. To eliminate the temporal chirp, we route the output field to the opposite side of the same CBG (rather than using a second CBG). We thus obtain a field in the form of a spatially resolved spectrum with a flat-phase wavefront. 

\subsection*{Fabrication of the large-area birefringent metasurface}
The birefringent metasurface of $14\times14$ mm$^2$ surface area exploited here represents the largest metasurface for shaping vectorial structured light to date. Such a device was fabricated using a process reliant on electron beam lithography and atomic layer deposition, as shown in Supplementary Fig. S9. The procedure is as follows: a fused silica substrate is first spin-coated with a positive tone electron beam resist (ZEP520A, Zeon SMI) that ultimately defines the height of the nanofins (800~nm). After baking the resist, the desired pillar patterns were written by exposing the resist using electron beam lithography (with an accelerating voltage of 150~kV), then developed in O-Xylene for 60~s. The developed pattern defines the geometry of the individual nanopillars. Afterward, TiO$_{2}$ was deposited via the atomic layer deposition process (ALD) to conformally fill the developed pattern. The excess layer of TiO$_{2}$ on top of the device was etched away using reactive ion etching to the original height of the resist. Finally, the resist was removed using a downstream ashing (oxygen radicals) process leaving the individual TiO$_{2}$ nanopillars surrounded by air. More details on the design and characterization of the birefringent metasurface are given in Supplementary Section II.D. 

\vspace{4mm}
\noindent
\textbf{Data availability statement}\\
The data that support the plots within this paper and other findings of this study are available from the corresponding author upon reasonable request.

\vspace{2mm}
\noindent
\textbf{Code availability statement}\\
The code to produce numerical results within this paper and other findings of this study are available from the corresponding author upon reasonable request.

\vspace{2mm}
\noindent
\textbf{Acknowledgments}\\
We thank A. Palmieri from Harvard University for providing the metasurface library and J. Keith Miller from Clemson University for providing the description of the diffractive phase plates. Work at UCF was supported by the U.S. Office of Naval Research (ONR) under award N00014-17-1-2458, and the ONR MURI program under award N00014-20-1-2789. Work at Harvard was supported by the ONR MURI program, under award N00014-20-1-2450, and by the Air Force Office of Scientific Research (AFOSR) under award FA9550-22-1-0243. Work at Clemson was supported by the ONR MURI program, under award N00014-20-1-2558. Work at Stanford was supported by the ONR MURI program, under award N00014-20-1-2450

\vspace{2mm}
\noindent
\textbf{Author contributions}\\
\noindent
MY and AFA developed the concept. MY designed the experiments, carried out the measurements, and analyzed the data, with assistance from LAH. The theoretical calculation has been performed by MY with the assistance of LAH and CG. The Supplementary document was prepared by MY and AFA with the assistance of AHD, EGJ, and CG. The polarization metasurface was designed by AHD and fabricated by JSP. The phase plates for the coordinate transformation were designed by JF and MY, and fabricated by JF. AFA, FC, EGJ, and SF supervised the research. All the authors contributed to writing the paper.

\noindent
Correspondence and requests for materials should be addressed to MY or AFA.\\(email: yessenov@ucf.edu; raddy@creol.ucf.edu)

\vspace{2mm}
\noindent
\textbf{Competing interests:} The authors declare no competing interests.

\clearpage
\bibliography{Meron}

\begin{thebibliography}{65}%
\makeatletter
\providecommand \@ifxundefined [1]{%
 \@ifx{#1\undefined}
}%
\providecommand \@ifnum [1]{%
 \ifnum #1\expandafter \@firstoftwo
 \else \expandafter \@secondoftwo
 \fi
}%
\providecommand \@ifx [1]{%
 \ifx #1\expandafter \@firstoftwo
 \else \expandafter \@secondoftwo
 \fi
}%
\providecommand \natexlab [1]{#1}%
\providecommand \enquote  [1]{``#1''}%
\providecommand \bibnamefont  [1]{#1}%
\providecommand \bibfnamefont [1]{#1}%
\providecommand \citenamefont [1]{#1}%
\providecommand \href@noop [0]{\@secondoftwo}%
\providecommand \href [0]{\begingroup \@sanitize@url \@href}%
\providecommand \@href[1]{\@@startlink{#1}\@@href}%
\providecommand \@@href[1]{\endgroup#1\@@endlink}%
\providecommand \@sanitize@url [0]{\catcode `\\12\catcode `\$12\catcode `\&12\catcode `\#12\catcode `\^12\catcode `\_12\catcode `\%12\relax}%
\providecommand \@@startlink[1]{}%
\providecommand \@@endlink[0]{}%
\providecommand \url  [0]{\begingroup\@sanitize@url \@url }%
\providecommand \@url [1]{\endgroup\@href {#1}{\urlprefix }}%
\providecommand \urlprefix  [0]{URL }%
\providecommand \Eprint [0]{\href }%
\providecommand \doibase [0]{http://dx.doi.org/}%
\providecommand \selectlanguage [0]{\@gobble}%
\providecommand \bibinfo  [0]{\@secondoftwo}%
\providecommand \bibfield  [0]{\@secondoftwo}%
\providecommand \translation [1]{[#1]}%
\providecommand \BibitemOpen [0]{}%
\providecommand \bibitemStop [0]{}%
\providecommand \bibitemNoStop [0]{.\EOS\space}%
\providecommand \EOS [0]{\spacefactor3000\relax}%
\providecommand \BibitemShut  [1]{\csname bibitem#1\endcsname}%
\let\auto@bib@innerbib\@empty
\bibitem [{\citenamefont {Skyrme}(1962)}]{Skyrme1962NucPhys}%
  \BibitemOpen
  \bibfield  {author} {\bibinfo {author} {\bibfnamefont {T.~H.~R.}\ \bibnamefont {Skyrme}},\ }\bibfield  {title} {\enquote {\bibinfo {title} {A unified field theory of mesons and baryons},}\ }\href@noop {} {\bibfield  {journal} {\bibinfo  {journal} {Nuclear Phys.}\ }\textbf {\bibinfo {volume} {31}},\ \bibinfo {pages} {556--569} (\bibinfo {year} {1962})}\BibitemShut {NoStop}%
\bibitem [{\citenamefont {Manton}\ and\ \citenamefont {Sutcliffe}(2004)}]{TopologicalSolitonBook}%
  \BibitemOpen
  \bibfield  {author} {\bibinfo {author} {\bibfnamefont {N.}~\bibnamefont {Manton}}\ and\ \bibinfo {author} {\bibfnamefont {P.}~\bibnamefont {Sutcliffe}},\ }\href@noop {} {\emph {\bibinfo {title} {Topological solitons}}}\ (\bibinfo  {publisher} {Cambridge University Press},\ \bibinfo {year} {2004})\BibitemShut {NoStop}%
\bibitem [{\citenamefont {Al~Khawaja}\ and\ \citenamefont {Stoof}(2001)}]{AlKhawaja2001}%
  \BibitemOpen
  \bibfield  {author} {\bibinfo {author} {\bibfnamefont {U.}~\bibnamefont {Al~Khawaja}}\ and\ \bibinfo {author} {\bibfnamefont {H.}~\bibnamefont {Stoof}},\ }\bibfield  {title} {\enquote {\bibinfo {title} {Skyrmions in a ferromagnetic {B}ose-{E}instein condensate},}\ }\href@noop {} {\bibfield  {journal} {\bibinfo  {journal} {Nature}\ }\textbf {\bibinfo {volume} {411}},\ \bibinfo {pages} {918--920} (\bibinfo {year} {2001})}\BibitemShut {NoStop}%
\bibitem [{\citenamefont {Han}(2017)}]{Han2017Book}%
  \BibitemOpen
  \bibfield  {author} {\bibinfo {author} {\bibfnamefont {J.~H.}\ \bibnamefont {Han}},\ }\href@noop {} {\emph {\bibinfo {title} {Skyrmions in condensed matter}}},\ Vol.\ \bibinfo {volume} {278}\ (\bibinfo  {publisher} {Springer},\ \bibinfo {year} {2017})\BibitemShut {NoStop}%
\bibitem [{\citenamefont {Duzgun}\ and\ \citenamefont {Nisoli}(2021)}]{PhysRevLett.126.047801}%
  \BibitemOpen
  \bibfield  {author} {\bibinfo {author} {\bibfnamefont {A.}~\bibnamefont {Duzgun}}\ and\ \bibinfo {author} {\bibfnamefont {C.}~\bibnamefont {Nisoli}},\ }\bibfield  {title} {\enquote {\bibinfo {title} {Skyrmion spin ice in liquid crystals},}\ }\href@noop {} {\bibfield  {journal} {\bibinfo  {journal} {Phys. Rev. Lett.}\ }\textbf {\bibinfo {volume} {126}},\ \bibinfo {pages} {047801} (\bibinfo {year} {2021})}\BibitemShut {NoStop}%
\bibitem [{\citenamefont {Bogdanov}\ and\ \citenamefont {Panagopoulos}(2020)}]{Bogdanov2020NatPhysRev}%
  \BibitemOpen
  \bibfield  {author} {\bibinfo {author} {\bibfnamefont {A.~N}\ \bibnamefont {Bogdanov}}\ and\ \bibinfo {author} {\bibfnamefont {C.}~\bibnamefont {Panagopoulos}},\ }\bibfield  {title} {\enquote {\bibinfo {title} {Physical foundations and basic properties of magnetic skyrmions},}\ }\href@noop {} {\bibfield  {journal} {\bibinfo  {journal} {Nat. Rev. Phys.}\ }\textbf {\bibinfo {volume} {2}},\ \bibinfo {pages} {492--498} (\bibinfo {year} {2020})}\BibitemShut {NoStop}%
\bibitem [{\citenamefont {Han}\ \emph {et~al.}(2022)\citenamefont {Han}, \citenamefont {Addiego}, \citenamefont {Prokhorenko}, \citenamefont {Wang}, \citenamefont {Fu}, \citenamefont {Nahas}, \citenamefont {Yan}, \citenamefont {Cai}, \citenamefont {Wei}, \citenamefont {Fang} \emph {et~al.}}]{Han2022}%
  \BibitemOpen
  \bibfield  {author} {\bibinfo {author} {\bibfnamefont {K.}~\bibnamefont {Han}}, \bibinfo {author} {\bibfnamefont {C.}~\bibnamefont {Addiego}}, \bibinfo {author} {\bibfnamefont {S.}~\bibnamefont {Prokhorenko}}, \bibinfo {author} {\bibfnamefont {M.}~\bibnamefont {Wang}}, \bibinfo {author} {\bibfnamefont {H.}~\bibnamefont {Fu}}, \bibinfo {author} {\bibfnamefont {Y.}~\bibnamefont {Nahas}}, \bibinfo {author} {\bibfnamefont {X.}~\bibnamefont {Yan}}, \bibinfo {author} {\bibfnamefont {S.}~\bibnamefont {Cai}}, \bibinfo {author} {\bibfnamefont {T.}~\bibnamefont {Wei}}, \bibinfo {author} {\bibfnamefont {Y.}~\bibnamefont {Fang}},  \emph {et~al.},\ }\bibfield  {title} {\enquote {\bibinfo {title} {High-density switchable skyrmion-like polar nanodomains integrated on silicon},}\ }\href@noop {} {\bibfield  {journal} {\bibinfo  {journal} {Nature}\ }\textbf {\bibinfo {volume} {603}},\ \bibinfo {pages} {63--67} (\bibinfo {year} {2022})}\BibitemShut {NoStop}%
\bibitem [{\citenamefont {Fert}\ \emph {et~al.}(2013)\citenamefont {Fert}, \citenamefont {Cros},\ and\ \citenamefont {Sampaio}}]{Fert2013NatNano}%
  \BibitemOpen
  \bibfield  {author} {\bibinfo {author} {\bibfnamefont {A.}~\bibnamefont {Fert}}, \bibinfo {author} {\bibfnamefont {V.}~\bibnamefont {Cros}}, \ and\ \bibinfo {author} {\bibfnamefont {J.}~\bibnamefont {Sampaio}},\ }\bibfield  {title} {\enquote {\bibinfo {title} {Skyrmions on the track},}\ }\href@noop {} {\bibfield  {journal} {\bibinfo  {journal} {Nat. Nanotech.}\ }\textbf {\bibinfo {volume} {8}},\ \bibinfo {pages} {152--156} (\bibinfo {year} {2013})}\BibitemShut {NoStop}%
\bibitem [{\citenamefont {Nagaosa}\ and\ \citenamefont {Tokura}(2013)}]{Nagaosa2013}%
  \BibitemOpen
  \bibfield  {author} {\bibinfo {author} {\bibfnamefont {N.}~\bibnamefont {Nagaosa}}\ and\ \bibinfo {author} {\bibfnamefont {Y.}~\bibnamefont {Tokura}},\ }\bibfield  {title} {\enquote {\bibinfo {title} {Topological properties and dynamics of magnetic skyrmions},}\ }\href@noop {} {\bibfield  {journal} {\bibinfo  {journal} {Nat. Nanotech.}\ }\textbf {\bibinfo {volume} {8}},\ \bibinfo {pages} {899--911} (\bibinfo {year} {2013})}\BibitemShut {NoStop}%
\bibitem [{\citenamefont {Fert}\ \emph {et~al.}(2017)\citenamefont {Fert}, \citenamefont {Reyren},\ and\ \citenamefont {Cros}}]{Fert2017NatRevMat}%
  \BibitemOpen
  \bibfield  {author} {\bibinfo {author} {\bibfnamefont {A.}~\bibnamefont {Fert}}, \bibinfo {author} {\bibfnamefont {N.}~\bibnamefont {Reyren}}, \ and\ \bibinfo {author} {\bibfnamefont {V.}~\bibnamefont {Cros}},\ }\bibfield  {title} {\enquote {\bibinfo {title} {Magnetic skyrmions: advances in physics and potential applications},}\ }\href@noop {} {\bibfield  {journal} {\bibinfo  {journal} {Nat. Rev. Mat.}\ }\textbf {\bibinfo {volume} {2}},\ \bibinfo {pages} {1--15} (\bibinfo {year} {2017})}\BibitemShut {NoStop}%
\bibitem [{\citenamefont {G{\"o}bel}\ \emph {et~al.}(2021)\citenamefont {G{\"o}bel}, \citenamefont {Mertig},\ and\ \citenamefont {Tretiakov}}]{Gobel2021Review}%
  \BibitemOpen
  \bibfield  {author} {\bibinfo {author} {\bibfnamefont {B.}~\bibnamefont {G{\"o}bel}}, \bibinfo {author} {\bibfnamefont {I.}~\bibnamefont {Mertig}}, \ and\ \bibinfo {author} {\bibfnamefont {O.~A.}\ \bibnamefont {Tretiakov}},\ }\bibfield  {title} {\enquote {\bibinfo {title} {Beyond skyrmions: Review and perspectives of alternative magnetic quasiparticles},}\ }\href@noop {} {\bibfield  {journal} {\bibinfo  {journal} {Phys. Rep.}\ }\textbf {\bibinfo {volume} {895}},\ \bibinfo {pages} {1--28} (\bibinfo {year} {2021})}\BibitemShut {NoStop}%
\bibitem [{\citenamefont {Cook}(2023)}]{Cook2023JoP}%
  \BibitemOpen
  \bibfield  {author} {\bibinfo {author} {\bibfnamefont {A.~M.}\ \bibnamefont {Cook}},\ }\bibfield  {title} {\enquote {\bibinfo {title} {Topological skyrmion phases of matter},}\ }\href@noop {} {\bibfield  {journal} {\bibinfo  {journal} {J. of Phys.: Cond. Matt.}\ }\textbf {\bibinfo {volume} {35}},\ \bibinfo {pages} {184001} (\bibinfo {year} {2023})}\BibitemShut {NoStop}%
\bibitem [{\citenamefont {Ackerman}\ and\ \citenamefont {Smalyukh}(2017)}]{Ackerman2017NatMat}%
  \BibitemOpen
  \bibfield  {author} {\bibinfo {author} {\bibfnamefont {P.~J.}\ \bibnamefont {Ackerman}}\ and\ \bibinfo {author} {\bibfnamefont {I.~I.}\ \bibnamefont {Smalyukh}},\ }\bibfield  {title} {\enquote {\bibinfo {title} {Static three-dimensional topological solitons in fluid chiral ferromagnets and colloids},}\ }\href@noop {} {\bibfield  {journal} {\bibinfo  {journal} {Nat. Mat.}\ }\textbf {\bibinfo {volume} {16}},\ \bibinfo {pages} {426--432} (\bibinfo {year} {2017})}\BibitemShut {NoStop}%
\bibitem [{\citenamefont {Lee}\ \emph {et~al.}(2018)\citenamefont {Lee}, \citenamefont {Gheorghe}, \citenamefont {Tiurev}, \citenamefont {Ollikainen}, \citenamefont {M{\"o}tt{\"o}nen},\ and\ \citenamefont {Hall}}]{Lee2018SciAdv}%
  \BibitemOpen
  \bibfield  {author} {\bibinfo {author} {\bibfnamefont {W.}~\bibnamefont {Lee}}, \bibinfo {author} {\bibfnamefont {A.~H.}\ \bibnamefont {Gheorghe}}, \bibinfo {author} {\bibfnamefont {K.}~\bibnamefont {Tiurev}}, \bibinfo {author} {\bibfnamefont {T.}~\bibnamefont {Ollikainen}}, \bibinfo {author} {\bibfnamefont {M.}~\bibnamefont {M{\"o}tt{\"o}nen}}, \ and\ \bibinfo {author} {\bibfnamefont {D.~S.}\ \bibnamefont {Hall}},\ }\bibfield  {title} {\enquote {\bibinfo {title} {Synthetic electromagnetic knot in a three-dimensional skyrmion},}\ }\href@noop {} {\bibfield  {journal} {\bibinfo  {journal} {Science Adv.}\ }\textbf {\bibinfo {volume} {4}},\ \bibinfo {pages} {eaao3820} (\bibinfo {year} {2018})}\BibitemShut {NoStop}%
\bibitem [{\citenamefont {Das}\ \emph {et~al.}(2019)\citenamefont {Das}, \citenamefont {Tang}, \citenamefont {Hong}, \citenamefont {Gon{\c{c}}alves}, \citenamefont {McCarter}, \citenamefont {Klewe}, \citenamefont {Nguyen}, \citenamefont {G{\'o}mez-Ortiz}, \citenamefont {Shafer}, \citenamefont {Arenholz} \emph {et~al.}}]{Das2019Nat}%
  \BibitemOpen
  \bibfield  {author} {\bibinfo {author} {\bibfnamefont {S.}~\bibnamefont {Das}}, \bibinfo {author} {\bibfnamefont {Y.~L.}\ \bibnamefont {Tang}}, \bibinfo {author} {\bibfnamefont {Z.}~\bibnamefont {Hong}}, \bibinfo {author} {\bibfnamefont {M.~A.~P.}\ \bibnamefont {Gon{\c{c}}alves}}, \bibinfo {author} {\bibfnamefont {M.~R.}\ \bibnamefont {McCarter}}, \bibinfo {author} {\bibfnamefont {C.}~\bibnamefont {Klewe}}, \bibinfo {author} {\bibfnamefont {K.~X}\ \bibnamefont {Nguyen}}, \bibinfo {author} {\bibfnamefont {F.}~\bibnamefont {G{\'o}mez-Ortiz}}, \bibinfo {author} {\bibfnamefont {P.}~\bibnamefont {Shafer}}, \bibinfo {author} {\bibfnamefont {E.}~\bibnamefont {Arenholz}},  \emph {et~al.},\ }\bibfield  {title} {\enquote {\bibinfo {title} {Observation of room-temperature polar skyrmions},}\ }\href@noop {} {\bibfield  {journal} {\bibinfo  {journal} {Nature}\ }\textbf {\bibinfo {volume} {568}},\ \bibinfo {pages} {368--372} (\bibinfo {year} {2019})}\BibitemShut {NoStop}%
\bibitem [{\citenamefont {Tai}\ and\ \citenamefont {Smalyukh}(2022)}]{Tai2022NC}%
  \BibitemOpen
  \bibfield  {author} {\bibinfo {author} {\bibfnamefont {J.}~\bibnamefont {Tai}, \bibfnamefont {J.~B .and~Wu}}\ and\ \bibinfo {author} {\bibfnamefont {I.~I}\ \bibnamefont {Smalyukh}},\ }\bibfield  {title} {\enquote {\bibinfo {title} {Geometric transformation and three-dimensional hopping of {H}opf solitons},}\ }\href@noop {} {\bibfield  {journal} {\bibinfo  {journal} {Nat. Comm.}\ }\textbf {\bibinfo {volume} {13}},\ \bibinfo {pages} {2986} (\bibinfo {year} {2022})}\BibitemShut {NoStop}%
\bibitem [{\citenamefont {Du}\ \emph {et~al.}(2019)\citenamefont {Du}, \citenamefont {Yang}, \citenamefont {Zayats},\ and\ \citenamefont {Yuan}}]{Du2019NP}%
  \BibitemOpen
  \bibfield  {author} {\bibinfo {author} {\bibfnamefont {L.}~\bibnamefont {Du}}, \bibinfo {author} {\bibfnamefont {A.}~\bibnamefont {Yang}}, \bibinfo {author} {\bibfnamefont {A.~V.}\ \bibnamefont {Zayats}}, \ and\ \bibinfo {author} {\bibfnamefont {X.}~\bibnamefont {Yuan}},\ }\bibfield  {title} {\enquote {\bibinfo {title} {Deep-subwavelength features of photonic skyrmions in a confined electromagnetic field with orbital angular momentum},}\ }\href@noop {} {\bibfield  {journal} {\bibinfo  {journal} {Nat. Phys.}\ }\textbf {\bibinfo {volume} {15}},\ \bibinfo {pages} {650--654} (\bibinfo {year} {2019})}\BibitemShut {NoStop}%
\bibitem [{\citenamefont {Ornelas}\ \emph {et~al.}(2024)\citenamefont {Ornelas}, \citenamefont {Nape}, \citenamefont {de~Mello~Koch},\ and\ \citenamefont {Forbes}}]{Ornelas2024NP}%
  \BibitemOpen
  \bibfield  {author} {\bibinfo {author} {\bibfnamefont {P.}~\bibnamefont {Ornelas}}, \bibinfo {author} {\bibfnamefont {I.}~\bibnamefont {Nape}}, \bibinfo {author} {\bibfnamefont {R.}~\bibnamefont {de~Mello~Koch}}, \ and\ \bibinfo {author} {\bibfnamefont {A.}~\bibnamefont {Forbes}},\ }\bibfield  {title} {\enquote {\bibinfo {title} {Non-local skyrmions as topologically resilient quantum entangled states of light},}\ }\href@noop {} {\bibfield  {journal} {\bibinfo  {journal} {Nat. Photon.}\ }\textbf {\bibinfo {volume} {18}},\ \bibinfo {pages} {258–266} (\bibinfo {year} {2024})}\BibitemShut {NoStop}%
\bibitem [{\citenamefont {Tsesses}\ \emph {et~al.}(2018)\citenamefont {Tsesses}, \citenamefont {Ostrovsky}, \citenamefont {Cohen}, \citenamefont {Gjonaj}, \citenamefont {Lindner},\ and\ \citenamefont {Bartal}}]{Tsesses2018Science}%
  \BibitemOpen
  \bibfield  {author} {\bibinfo {author} {\bibfnamefont {S.}~\bibnamefont {Tsesses}}, \bibinfo {author} {\bibfnamefont {E.}~\bibnamefont {Ostrovsky}}, \bibinfo {author} {\bibfnamefont {K.}~\bibnamefont {Cohen}}, \bibinfo {author} {\bibfnamefont {B.}~\bibnamefont {Gjonaj}}, \bibinfo {author} {\bibfnamefont {N.~H}\ \bibnamefont {Lindner}}, \ and\ \bibinfo {author} {\bibfnamefont {G.}~\bibnamefont {Bartal}},\ }\bibfield  {title} {\enquote {\bibinfo {title} {Optical skyrmion lattice in evanescent electromagnetic fields},}\ }\href@noop {} {\bibfield  {journal} {\bibinfo  {journal} {Science}\ }\textbf {\bibinfo {volume} {361}},\ \bibinfo {pages} {993--996} (\bibinfo {year} {2018})}\BibitemShut {NoStop}%
\bibitem [{\citenamefont {Dai}\ \emph {et~al.}(2020)\citenamefont {Dai}, \citenamefont {Zhou}, \citenamefont {Ghosh}, \citenamefont {Mong}, \citenamefont {Kubo}, \citenamefont {Huang},\ and\ \citenamefont {Petek}}]{Dai2020Nat}%
  \BibitemOpen
  \bibfield  {author} {\bibinfo {author} {\bibfnamefont {Y.}~\bibnamefont {Dai}}, \bibinfo {author} {\bibfnamefont {Z.}~\bibnamefont {Zhou}}, \bibinfo {author} {\bibfnamefont {A.}~\bibnamefont {Ghosh}}, \bibinfo {author} {\bibfnamefont {R.~S.~K.}\ \bibnamefont {Mong}}, \bibinfo {author} {\bibfnamefont {A.}~\bibnamefont {Kubo}}, \bibinfo {author} {\bibfnamefont {C.}~\bibnamefont {Huang}}, \ and\ \bibinfo {author} {\bibfnamefont {H.}~\bibnamefont {Petek}},\ }\bibfield  {title} {\enquote {\bibinfo {title} {Plasmonic topological quasiparticle on the nanometre and femtosecond scales},}\ }\href@noop {} {\bibfield  {journal} {\bibinfo  {journal} {Nature}\ }\textbf {\bibinfo {volume} {588}},\ \bibinfo {pages} {616--619} (\bibinfo {year} {2020})}\BibitemShut {NoStop}%
\bibitem [{\citenamefont {Gao}\ \emph {et~al.}(2020)\citenamefont {Gao}, \citenamefont {Speirits}, \citenamefont {Castellucci}, \citenamefont {Franke-Arnold}, \citenamefont {Barnett},\ and\ \citenamefont {G\"otte}}]{Gao2020PRA}%
  \BibitemOpen
  \bibfield  {author} {\bibinfo {author} {\bibfnamefont {S.}~\bibnamefont {Gao}}, \bibinfo {author} {\bibfnamefont {F.~C.}\ \bibnamefont {Speirits}}, \bibinfo {author} {\bibfnamefont {F.}~\bibnamefont {Castellucci}}, \bibinfo {author} {\bibfnamefont {S.}~\bibnamefont {Franke-Arnold}}, \bibinfo {author} {\bibfnamefont {S.~M.}\ \bibnamefont {Barnett}}, \ and\ \bibinfo {author} {\bibfnamefont {J.~B.}\ \bibnamefont {G\"otte}},\ }\bibfield  {title} {\enquote {\bibinfo {title} {Paraxial skyrmionic beams},}\ }\href@noop {} {\bibfield  {journal} {\bibinfo  {journal} {Phys. Rev. A}\ }\textbf {\bibinfo {volume} {102}},\ \bibinfo {pages} {053513} (\bibinfo {year} {2020})}\BibitemShut {NoStop}%
\bibitem [{\citenamefont {Shen}\ \emph {et~al.}(2022)\citenamefont {Shen}, \citenamefont {Mart{\'\i}nez},\ and\ \citenamefont {Rosales-Guzm{\'a}n}}]{Shen2022ACS}%
  \BibitemOpen
  \bibfield  {author} {\bibinfo {author} {\bibfnamefont {Y.}~\bibnamefont {Shen}}, \bibinfo {author} {\bibfnamefont {E.~C.}\ \bibnamefont {Mart{\'\i}nez}}, \ and\ \bibinfo {author} {\bibfnamefont {C.}~\bibnamefont {Rosales-Guzm{\'a}n}},\ }\bibfield  {title} {\enquote {\bibinfo {title} {Generation of optical skyrmions with tunable topological textures},}\ }\href@noop {} {\bibfield  {journal} {\bibinfo  {journal} {ACS Photon.}\ }\textbf {\bibinfo {volume} {9}},\ \bibinfo {pages} {296--303} (\bibinfo {year} {2022})}\BibitemShut {NoStop}%
\bibitem [{\citenamefont {Marco}\ \emph {et~al.}(2024)\citenamefont {Marco}, \citenamefont {Herrera}, \citenamefont {Brasselet},\ and\ \citenamefont {Alonso}}]{Marco2024ACSPhot}%
  \BibitemOpen
  \bibfield  {author} {\bibinfo {author} {\bibfnamefont {D.}~\bibnamefont {Marco}}, \bibinfo {author} {\bibfnamefont {I.}~\bibnamefont {Herrera}}, \bibinfo {author} {\bibfnamefont {S.}~\bibnamefont {Brasselet}}, \ and\ \bibinfo {author} {\bibfnamefont {M.~A.}\ \bibnamefont {Alonso}},\ }\bibfield  {title} {\enquote {\bibinfo {title} {Propagation-invariant optical meron lattices},}\ }\href@noop {} {\bibfield  {journal} {\bibinfo  {journal} {ACS Photon.}\ } (\bibinfo {year} {2024})}\BibitemShut {NoStop}%
\bibitem [{\citenamefont {Sugic}\ \emph {et~al.}(2021)\citenamefont {Sugic}, \citenamefont {Droop}, \citenamefont {Otte}, \citenamefont {Ehrmanntraut}, \citenamefont {Nori}, \citenamefont {Ruostekoski}, \citenamefont {Denz},\ and\ \citenamefont {Dennis}}]{Sugic2021NC}%
  \BibitemOpen
  \bibfield  {author} {\bibinfo {author} {\bibfnamefont {D.}~\bibnamefont {Sugic}}, \bibinfo {author} {\bibfnamefont {R.}~\bibnamefont {Droop}}, \bibinfo {author} {\bibfnamefont {E.}~\bibnamefont {Otte}}, \bibinfo {author} {\bibfnamefont {D.}~\bibnamefont {Ehrmanntraut}}, \bibinfo {author} {\bibfnamefont {F.}~\bibnamefont {Nori}}, \bibinfo {author} {\bibfnamefont {J.}~\bibnamefont {Ruostekoski}}, \bibinfo {author} {\bibfnamefont {C.}~\bibnamefont {Denz}}, \ and\ \bibinfo {author} {\bibfnamefont {M.~R.}\ \bibnamefont {Dennis}},\ }\bibfield  {title} {\enquote {\bibinfo {title} {Particle-like topologies in light},}\ }\href@noop {} {\bibfield  {journal} {\bibinfo  {journal} {Nat. Comm.}\ }\textbf {\bibinfo {volume} {12}},\ \bibinfo {pages} {6785} (\bibinfo {year} {2021})}\BibitemShut {NoStop}%
\bibitem [{\citenamefont {Wang}\ and\ \citenamefont {Fan}(2023)}]{Wang2023PRL}%
  \BibitemOpen
  \bibfield  {author} {\bibinfo {author} {\bibfnamefont {H.}~\bibnamefont {Wang}}\ and\ \bibinfo {author} {\bibfnamefont {S.}~\bibnamefont {Fan}},\ }\bibfield  {title} {\enquote {\bibinfo {title} {Photonic spin hopfions and monopole loops},}\ }\href@noop {} {\bibfield  {journal} {\bibinfo  {journal} {Phys. Rev. Lett.}\ }\textbf {\bibinfo {volume} {131}},\ \bibinfo {pages} {263801} (\bibinfo {year} {2023})}\BibitemShut {NoStop}%
\bibitem [{\citenamefont {Yessenov}\ \emph {et~al.}(2022{\natexlab{a}})\citenamefont {Yessenov}, \citenamefont {Hall}, \citenamefont {Schepler},\ and\ \citenamefont {Abouraddy}}]{Yessenov22AOP}%
  \BibitemOpen
  \bibfield  {author} {\bibinfo {author} {\bibfnamefont {M.}~\bibnamefont {Yessenov}}, \bibinfo {author} {\bibfnamefont {A.~A.}\ \bibnamefont {Hall}}, \bibinfo {author} {\bibfnamefont {K.~L.}\ \bibnamefont {Schepler}}, \ and\ \bibinfo {author} {\bibfnamefont {A.~F.}\ \bibnamefont {Abouraddy}},\ }\bibfield  {title} {\enquote {\bibinfo {title} {Space-time wave packets},}\ }\href@noop {} {\bibfield  {journal} {\bibinfo  {journal} {Adv. Opt. Photon.}\ }\textbf {\bibinfo {volume} {14}},\ \bibinfo {pages} {455--570} (\bibinfo {year} {2022}{\natexlab{a}})}\BibitemShut {NoStop}%
\bibitem [{\citenamefont {Shen}\ \emph {et~al.}(2023)\citenamefont {Shen}, \citenamefont {Zhan}, \citenamefont {Wright}, \citenamefont {Christodoulides}, \citenamefont {Wise}, \citenamefont {Willner}, \citenamefont {Zou}, \citenamefont {Zhao}, \citenamefont {Porras}, \citenamefont {Chong} \emph {et~al.}}]{shen2023JoO}%
  \BibitemOpen
  \bibfield  {author} {\bibinfo {author} {\bibfnamefont {Y.}~\bibnamefont {Shen}}, \bibinfo {author} {\bibfnamefont {Q.}~\bibnamefont {Zhan}}, \bibinfo {author} {\bibfnamefont {L.~G.}\ \bibnamefont {Wright}}, \bibinfo {author} {\bibfnamefont {D.~N.}\ \bibnamefont {Christodoulides}}, \bibinfo {author} {\bibfnamefont {F.~W.}\ \bibnamefont {Wise}}, \bibinfo {author} {\bibfnamefont {A.~E.}\ \bibnamefont {Willner}}, \bibinfo {author} {\bibfnamefont {K.}~\bibnamefont {Zou}}, \bibinfo {author} {\bibfnamefont {Z.}~\bibnamefont {Zhao}}, \bibinfo {author} {\bibfnamefont {A.~A.}\ \bibnamefont {Porras}}, \bibinfo {author} {\bibfnamefont {A.}~\bibnamefont {Chong}},  \emph {et~al.},\ }\bibfield  {title} {\enquote {\bibinfo {title} {Roadmap on spatiotemporal light fields},}\ }\href@noop {} {\bibfield  {journal} {\bibinfo  {journal} {J. of Opt.}\ }\textbf {\bibinfo {volume} {25}},\ \bibinfo {pages} {093001} (\bibinfo {year} {2023})}\BibitemShut {NoStop}%
\bibitem [{\citenamefont {Guo}\ \emph {et~al.}(2020)\citenamefont {Guo}, \citenamefont {Xiao}, \citenamefont {Guo}, \citenamefont {Yuan},\ and\ \citenamefont {Fan}}]{Guo2020PRL}%
  \BibitemOpen
  \bibfield  {author} {\bibinfo {author} {\bibfnamefont {C.}~\bibnamefont {Guo}}, \bibinfo {author} {\bibfnamefont {M.}~\bibnamefont {Xiao}}, \bibinfo {author} {\bibfnamefont {Y.}~\bibnamefont {Guo}}, \bibinfo {author} {\bibfnamefont {L.}~\bibnamefont {Yuan}}, \ and\ \bibinfo {author} {\bibfnamefont {S.}~\bibnamefont {Fan}},\ }\bibfield  {title} {\enquote {\bibinfo {title} {Meron spin textures in momentum space},}\ }\href@noop {} {\bibfield  {journal} {\bibinfo  {journal} {Phys. Rev. Lett.}\ }\textbf {\bibinfo {volume} {124}},\ \bibinfo {pages} {106103} (\bibinfo {year} {2020})}\BibitemShut {NoStop}%
\bibitem [{\citenamefont {Yessenov}\ \emph {et~al.}(2022{\natexlab{b}})\citenamefont {Yessenov}, \citenamefont {Free}, \citenamefont {Chen}, \citenamefont {Johnson}, \citenamefont {Lavery}, \citenamefont {Alonso},\ and\ \citenamefont {Abouraddy}}]{Yessenov22NC}%
  \BibitemOpen
  \bibfield  {author} {\bibinfo {author} {\bibfnamefont {M.}~\bibnamefont {Yessenov}}, \bibinfo {author} {\bibfnamefont {J.}~\bibnamefont {Free}}, \bibinfo {author} {\bibfnamefont {Z.}~\bibnamefont {Chen}}, \bibinfo {author} {\bibfnamefont {E.~G.}\ \bibnamefont {Johnson}}, \bibinfo {author} {\bibfnamefont {M.~P.~J.}\ \bibnamefont {Lavery}}, \bibinfo {author} {\bibfnamefont {M.~A.}\ \bibnamefont {Alonso}}, \ and\ \bibinfo {author} {\bibfnamefont {A.~F.}\ \bibnamefont {Abouraddy}},\ }\bibfield  {title} {\enquote {\bibinfo {title} {Space-time wave packets localized in all dimensions},}\ }\href@noop {} {\bibfield  {journal} {\bibinfo  {journal} {Nat. Commun.}\ }\textbf {\bibinfo {volume} {13}},\ \bibinfo {pages} {4573} (\bibinfo {year} {2022}{\natexlab{b}})}\BibitemShut {NoStop}%
\bibitem [{\citenamefont {Yessenov}\ \emph {et~al.}(2022{\natexlab{c}})\citenamefont {Yessenov}, \citenamefont {Chen}, \citenamefont {Lavery},\ and\ \citenamefont {Abouraddy}}]{Yessenov22OL}%
  \BibitemOpen
  \bibfield  {author} {\bibinfo {author} {\bibfnamefont {M.}~\bibnamefont {Yessenov}}, \bibinfo {author} {\bibfnamefont {Z.}~\bibnamefont {Chen}}, \bibinfo {author} {\bibfnamefont {M.~P.~J.}\ \bibnamefont {Lavery}}, \ and\ \bibinfo {author} {\bibfnamefont {A.~F.}\ \bibnamefont {Abouraddy}},\ }\bibfield  {title} {\enquote {\bibinfo {title} {Vector space-time wave packets},}\ }\href@noop {} {\bibfield  {journal} {\bibinfo  {journal} {Opt. Lett.}\ }\textbf {\bibinfo {volume} {47}},\ \bibinfo {pages} {4131--4134} (\bibinfo {year} {2022}{\natexlab{c}})}\BibitemShut {NoStop}%
\bibitem [{\citenamefont {Guo}\ \emph {et~al.}(2021)\citenamefont {Guo}, \citenamefont {Xiao}, \citenamefont {Orenstein},\ and\ \citenamefont {Fan}}]{Guo21Light}%
  \BibitemOpen
  \bibfield  {author} {\bibinfo {author} {\bibfnamefont {C.}~\bibnamefont {Guo}}, \bibinfo {author} {\bibfnamefont {M.}~\bibnamefont {Xiao}}, \bibinfo {author} {\bibfnamefont {M.}~\bibnamefont {Orenstein}}, \ and\ \bibinfo {author} {\bibfnamefont {S.}~\bibnamefont {Fan}},\ }\bibfield  {title} {\enquote {\bibinfo {title} {Structured {3D} linear space–time light bullets by nonlocal nanophotonics},}\ }\href@noop {} {\bibfield  {journal} {\bibinfo  {journal} {Light Sci. \& Appl.}\ }\textbf {\bibinfo {volume} {10}},\ \bibinfo {pages} {160} (\bibinfo {year} {2021})}\BibitemShut {NoStop}%
\bibitem [{\citenamefont {Loder}\ \emph {et~al.}(2017)\citenamefont {Loder}, \citenamefont {Kampf}, \citenamefont {Kopp},\ and\ \citenamefont {Braak}}]{Loder2017PRB}%
  \BibitemOpen
  \bibfield  {author} {\bibinfo {author} {\bibfnamefont {F.}~\bibnamefont {Loder}}, \bibinfo {author} {\bibfnamefont {A.~P.}\ \bibnamefont {Kampf}}, \bibinfo {author} {\bibfnamefont {T.}~\bibnamefont {Kopp}}, \ and\ \bibinfo {author} {\bibfnamefont {D.}~\bibnamefont {Braak}},\ }\bibfield  {title} {\enquote {\bibinfo {title} {Momentum-space spin texture in a topological superconductor},}\ }\href@noop {} {\bibfield  {journal} {\bibinfo  {journal} {Phys. Rev. B}\ }\textbf {\bibinfo {volume} {96}},\ \bibinfo {pages} {024508} (\bibinfo {year} {2017})}\BibitemShut {NoStop}%
\bibitem [{\citenamefont {Kondakci}\ and\ \citenamefont {Abouraddy}(2019)}]{Kondakci19NC}%
  \BibitemOpen
  \bibfield  {author} {\bibinfo {author} {\bibfnamefont {H.~E.}\ \bibnamefont {Kondakci}}\ and\ \bibinfo {author} {\bibfnamefont {A.~F.}\ \bibnamefont {Abouraddy}},\ }\bibfield  {title} {\enquote {\bibinfo {title} {Optical space-time wave packets of arbitrary group velocity in free space},}\ }\href@noop {} {\bibfield  {journal} {\bibinfo  {journal} {Nat. Commun.}\ }\textbf {\bibinfo {volume} {10}},\ \bibinfo {pages} {929} (\bibinfo {year} {2019})}\BibitemShut {NoStop}%
\bibitem [{\citenamefont {Bhaduri}\ \emph {et~al.}(2020)\citenamefont {Bhaduri}, \citenamefont {Yessenov},\ and\ \citenamefont {Abouraddy}}]{Bhaduri20NP}%
  \BibitemOpen
  \bibfield  {author} {\bibinfo {author} {\bibfnamefont {B.}~\bibnamefont {Bhaduri}}, \bibinfo {author} {\bibfnamefont {M.}~\bibnamefont {Yessenov}}, \ and\ \bibinfo {author} {\bibfnamefont {A.~F.}\ \bibnamefont {Abouraddy}},\ }\bibfield  {title} {\enquote {\bibinfo {title} {Anomalous refraction of optical space-time wave packets},}\ }\href@noop {} {\bibfield  {journal} {\bibinfo  {journal} {Nat. Photon.}\ }\textbf {\bibinfo {volume} {14}},\ \bibinfo {pages} {416--421} (\bibinfo {year} {2020})}\BibitemShut {NoStop}%
\bibitem [{\citenamefont {Kondakci}\ and\ \citenamefont {Abouraddy}(2018)}]{Kondakci18OL}%
  \BibitemOpen
  \bibfield  {author} {\bibinfo {author} {\bibfnamefont {H.~E.}\ \bibnamefont {Kondakci}}\ and\ \bibinfo {author} {\bibfnamefont {A.~F.}\ \bibnamefont {Abouraddy}},\ }\bibfield  {title} {\enquote {\bibinfo {title} {Self-healing of space-time light sheets},}\ }\href@noop {} {\bibfield  {journal} {\bibinfo  {journal} {Opt. Lett.}\ }\textbf {\bibinfo {volume} {43}},\ \bibinfo {pages} {3830--3833} (\bibinfo {year} {2018})}\BibitemShut {NoStop}%
\bibitem [{\citenamefont {Gao}\ \emph {et~al.}(2018)\citenamefont {Gao}, \citenamefont {Xue}, \citenamefont {Yang}, \citenamefont {Lai}, \citenamefont {Yu}, \citenamefont {Lin}, \citenamefont {Chong}, \citenamefont {Shvets},\ and\ \citenamefont {Zhang}}]{Gao2018NP}%
  \BibitemOpen
  \bibfield  {author} {\bibinfo {author} {\bibfnamefont {F.}~\bibnamefont {Gao}}, \bibinfo {author} {\bibfnamefont {H.}~\bibnamefont {Xue}}, \bibinfo {author} {\bibfnamefont {Z.}~\bibnamefont {Yang}}, \bibinfo {author} {\bibfnamefont {K.}~\bibnamefont {Lai}}, \bibinfo {author} {\bibfnamefont {Y.}~\bibnamefont {Yu}}, \bibinfo {author} {\bibfnamefont {X.}~\bibnamefont {Lin}}, \bibinfo {author} {\bibfnamefont {Y.}~\bibnamefont {Chong}}, \bibinfo {author} {\bibfnamefont {G.}~\bibnamefont {Shvets}}, \ and\ \bibinfo {author} {\bibfnamefont {B.}~\bibnamefont {Zhang}},\ }\bibfield  {title} {\enquote {\bibinfo {title} {Topologically protected refraction of robust kink states in valley photonic crystals},}\ }\href@noop {} {\bibfield  {journal} {\bibinfo  {journal} {Nat. Phys.}\ }\textbf {\bibinfo {volume} {14}},\ \bibinfo {pages} {140--144} (\bibinfo {year} {2018})}\BibitemShut {NoStop}%
\bibitem [{\citenamefont {Yang}\ \emph {et~al.}(2020)\citenamefont {Yang}, \citenamefont {Yamagami}, \citenamefont {Yu}, \citenamefont {Pitchappa}, \citenamefont {Webber}, \citenamefont {Zhang}, \citenamefont {Fujita}, \citenamefont {Nagatsuma},\ and\ \citenamefont {Singh}}]{Yang2020NP}%
  \BibitemOpen
  \bibfield  {author} {\bibinfo {author} {\bibfnamefont {Y.}~\bibnamefont {Yang}}, \bibinfo {author} {\bibfnamefont {Y.}~\bibnamefont {Yamagami}}, \bibinfo {author} {\bibfnamefont {X.}~\bibnamefont {Yu}}, \bibinfo {author} {\bibfnamefont {P.}~\bibnamefont {Pitchappa}}, \bibinfo {author} {\bibfnamefont {J.}~\bibnamefont {Webber}}, \bibinfo {author} {\bibfnamefont {B.}~\bibnamefont {Zhang}}, \bibinfo {author} {\bibfnamefont {M.}~\bibnamefont {Fujita}}, \bibinfo {author} {\bibfnamefont {T.}~\bibnamefont {Nagatsuma}}, \ and\ \bibinfo {author} {\bibfnamefont {R.}~\bibnamefont {Singh}},\ }\bibfield  {title} {\enquote {\bibinfo {title} {Terahertz topological photonics for on-chip communication},}\ }\href@noop {} {\bibfield  {journal} {\bibinfo  {journal} {Nat. Photon.}\ }\textbf {\bibinfo {volume} {14}},\ \bibinfo {pages} {446--451} (\bibinfo {year} {2020})}\BibitemShut {NoStop}%
\bibitem [{\citenamefont {Rivera}\ and\ \citenamefont {Kaminer}(2020)}]{Rivera2020NatRevPhys}%
  \BibitemOpen
  \bibfield  {author} {\bibinfo {author} {\bibfnamefont {N.}~\bibnamefont {Rivera}}\ and\ \bibinfo {author} {\bibfnamefont {I.}~\bibnamefont {Kaminer}},\ }\bibfield  {title} {\enquote {\bibinfo {title} {Light--matter interactions with photonic quasiparticles},}\ }\href@noop {} {\bibfield  {journal} {\bibinfo  {journal} {Nat. Rev. Phys.}\ }\textbf {\bibinfo {volume} {2}},\ \bibinfo {pages} {538--561} (\bibinfo {year} {2020})}\BibitemShut {NoStop}%
\bibitem [{\citenamefont {Beckley}\ \emph {et~al.}(2010)\citenamefont {Beckley}, \citenamefont {Brown},\ and\ \citenamefont {Alonso}}]{Beckley2010OE}%
  \BibitemOpen
  \bibfield  {author} {\bibinfo {author} {\bibfnamefont {A.~M.}\ \bibnamefont {Beckley}}, \bibinfo {author} {\bibfnamefont {T.~G.}\ \bibnamefont {Brown}}, \ and\ \bibinfo {author} {\bibfnamefont {M.~A.}\ \bibnamefont {Alonso}},\ }\bibfield  {title} {\enquote {\bibinfo {title} {Full {P}oincar{\'e} beams},}\ }\href@noop {} {\bibfield  {journal} {\bibinfo  {journal} {Opt. Express}\ }\textbf {\bibinfo {volume} {18}},\ \bibinfo {pages} {10777--10785} (\bibinfo {year} {2010})}\BibitemShut {NoStop}%
\bibitem [{\citenamefont {Shen}\ \emph {et~al.}(2024)\citenamefont {Shen}, \citenamefont {Zhang}, \citenamefont {Shi}, \citenamefont {Du}, \citenamefont {Yuan},\ and\ \citenamefont {Zayats}}]{Shen2023NP}%
  \BibitemOpen
  \bibfield  {author} {\bibinfo {author} {\bibfnamefont {Y.}~\bibnamefont {Shen}}, \bibinfo {author} {\bibfnamefont {Q.}~\bibnamefont {Zhang}}, \bibinfo {author} {\bibfnamefont {P.}~\bibnamefont {Shi}}, \bibinfo {author} {\bibfnamefont {L.}~\bibnamefont {Du}}, \bibinfo {author} {\bibfnamefont {X.}~\bibnamefont {Yuan}}, \ and\ \bibinfo {author} {\bibfnamefont {A.~V.}\ \bibnamefont {Zayats}},\ }\bibfield  {title} {\enquote {\bibinfo {title} {Optical skyrmions and other topological quasiparticles of light},}\ }\href@noop {} {\bibfield  {journal} {\bibinfo  {journal} {Nat. Photon.}\ }\textbf {\bibinfo {volume} {18}},\ \bibinfo {pages} {15--25} (\bibinfo {year} {2024})}\BibitemShut {NoStop}%
\bibitem [{\citenamefont {Kondakci}\ and\ \citenamefont {Abouraddy}(2017)}]{Kondakci17NP}%
  \BibitemOpen
  \bibfield  {author} {\bibinfo {author} {\bibfnamefont {H.~E.}\ \bibnamefont {Kondakci}}\ and\ \bibinfo {author} {\bibfnamefont {A.~F.}\ \bibnamefont {Abouraddy}},\ }\bibfield  {title} {\enquote {\bibinfo {title} {Diffraction-free space-time beams},}\ }\href@noop {} {\bibfield  {journal} {\bibinfo  {journal} {Nat. Photon.}\ }\textbf {\bibinfo {volume} {11}},\ \bibinfo {pages} {733--740} (\bibinfo {year} {2017})}\BibitemShut {NoStop}%
\bibitem [{\citenamefont {Kaim}\ \emph {et~al.}(2014)\citenamefont {Kaim}, \citenamefont {Mokhov}, \citenamefont {Zeldovich},\ and\ \citenamefont {Glebov}}]{Kaim13SPIE}%
  \BibitemOpen
  \bibfield  {author} {\bibinfo {author} {\bibfnamefont {S.}~\bibnamefont {Kaim}}, \bibinfo {author} {\bibfnamefont {S.}~\bibnamefont {Mokhov}}, \bibinfo {author} {\bibfnamefont {B.~Y.}\ \bibnamefont {Zeldovich}}, \ and\ \bibinfo {author} {\bibfnamefont {L.~B.}\ \bibnamefont {Glebov}},\ }\bibfield  {title} {\enquote {\bibinfo {title} {Stretching and compressing of short laser pulses by chirped volume {B}ragg gratings: analytic and numerical modeling},}\ }\href@noop {} {\bibfield  {journal} {\bibinfo  {journal} {Opt. Eng.}\ }\textbf {\bibinfo {volume} {53}},\ \bibinfo {pages} {051509} (\bibinfo {year} {2014})}\BibitemShut {NoStop}%
\bibitem [{\citenamefont {Bryngdahl}(1974)}]{Bryngdahl74JOSA}%
  \BibitemOpen
  \bibfield  {author} {\bibinfo {author} {\bibfnamefont {O.}~\bibnamefont {Bryngdahl}},\ }\bibfield  {title} {\enquote {\bibinfo {title} {Geometrical transformations in optics},}\ }\href@noop {} {\bibfield  {journal} {\bibinfo  {journal} {J. Opt. Soc. Am.}\ }\textbf {\bibinfo {volume} {64}},\ \bibinfo {pages} {1092--1099} (\bibinfo {year} {1974})}\BibitemShut {NoStop}%
\bibitem [{\citenamefont {Hossack}\ \emph {et~al.}(1987)\citenamefont {Hossack}, \citenamefont {Darling},\ and\ \citenamefont {Dahdouh}}]{Hossack87JOMO}%
  \BibitemOpen
  \bibfield  {author} {\bibinfo {author} {\bibfnamefont {W.~J.}\ \bibnamefont {Hossack}}, \bibinfo {author} {\bibfnamefont {A.~M.}\ \bibnamefont {Darling}}, \ and\ \bibinfo {author} {\bibfnamefont {A.}~\bibnamefont {Dahdouh}},\ }\bibfield  {title} {\enquote {\bibinfo {title} {Coordinate transformations with multiple computer-generated optical elements},}\ }\href@noop {} {\bibfield  {journal} {\bibinfo  {journal} {J. Mod. Opt.}\ }\textbf {\bibinfo {volume} {34}},\ \bibinfo {pages} {1235--1250} (\bibinfo {year} {1987})}\BibitemShut {NoStop}%
\bibitem [{\citenamefont {Berkhout}\ \emph {et~al.}(2010)\citenamefont {Berkhout}, \citenamefont {Lavery}, \citenamefont {Courtial}, \citenamefont {Beijersbergen},\ and\ \citenamefont {Padgett}}]{Berkhout10PRL}%
  \BibitemOpen
  \bibfield  {author} {\bibinfo {author} {\bibfnamefont {G.~C.~G}\ \bibnamefont {Berkhout}}, \bibinfo {author} {\bibfnamefont {M.~P.~J.}\ \bibnamefont {Lavery}}, \bibinfo {author} {\bibfnamefont {J.}~\bibnamefont {Courtial}}, \bibinfo {author} {\bibfnamefont {M.~W.}\ \bibnamefont {Beijersbergen}}, \ and\ \bibinfo {author} {\bibfnamefont {M.~J.}\ \bibnamefont {Padgett}},\ }\bibfield  {title} {\enquote {\bibinfo {title} {Efficient sorting of orbital angular momentum states of light},}\ }\href@noop {} {\bibfield  {journal} {\bibinfo  {journal} {Phys. Rev. Let.}\ }\textbf {\bibinfo {volume} {105}},\ \bibinfo {pages} {153601} (\bibinfo {year} {2010})}\BibitemShut {NoStop}%
\bibitem [{\citenamefont {Sung}\ \emph {et~al.}(2005)\citenamefont {Sung}, \citenamefont {Hockel},\ and\ \citenamefont {Johnson}}]{Sung2005OL}%
  \BibitemOpen
  \bibfield  {author} {\bibinfo {author} {\bibfnamefont {J.}~\bibnamefont {Sung}}, \bibinfo {author} {\bibfnamefont {H.}~\bibnamefont {Hockel}}, \ and\ \bibinfo {author} {\bibfnamefont {E.~G}\ \bibnamefont {Johnson}},\ }\bibfield  {title} {\enquote {\bibinfo {title} {Analog micro-optics fabrication by use of a two-dimensional binary phase-grating mask},}\ }\href@noop {} {\bibfield  {journal} {\bibinfo  {journal} {Opt. Lett.}\ }\textbf {\bibinfo {volume} {30}},\ \bibinfo {pages} {150--152} (\bibinfo {year} {2005})}\BibitemShut {NoStop}%
\bibitem [{\citenamefont {Rubin}\ \emph {et~al.}(2021)\citenamefont {Rubin}, \citenamefont {Shi},\ and\ \citenamefont {Capasso}}]{Rubin21AOP}%
  \BibitemOpen
  \bibfield  {author} {\bibinfo {author} {\bibfnamefont {N.~A.}\ \bibnamefont {Rubin}}, \bibinfo {author} {\bibfnamefont {Z.}~\bibnamefont {Shi}}, \ and\ \bibinfo {author} {\bibfnamefont {F.}~\bibnamefont {Capasso}},\ }\bibfield  {title} {\enquote {\bibinfo {title} {Polarization in diffractive optics and metasurfaces},}\ }\href@noop {} {\bibfield  {journal} {\bibinfo  {journal} {Adv. Opt. Photon.}\ }\textbf {\bibinfo {volume} {13}},\ \bibinfo {pages} {836--970} (\bibinfo {year} {2021})}\BibitemShut {NoStop}%
\bibitem [{\citenamefont {Dorrah}\ and\ \citenamefont {Capasso}(2022)}]{Dorrah22ScienceRev}%
  \BibitemOpen
  \bibfield  {author} {\bibinfo {author} {\bibfnamefont {A.~H.}\ \bibnamefont {Dorrah}}\ and\ \bibinfo {author} {\bibfnamefont {F.}~\bibnamefont {Capasso}},\ }\bibfield  {title} {\enquote {\bibinfo {title} {Tunable structured light with flat optics},}\ }\href@noop {} {\bibfield  {journal} {\bibinfo  {journal} {Science}\ }\textbf {\bibinfo {volume} {376}},\ \bibinfo {pages} {eabi6860} (\bibinfo {year} {2022})}\BibitemShut {NoStop}%
\bibitem [{\citenamefont {Yu}\ \emph {et~al.}(2011)\citenamefont {Yu}, \citenamefont {Genevet}, \citenamefont {Kats}, \citenamefont {Aieta}, \citenamefont {Tetienne}, \citenamefont {Capasso},\ and\ \citenamefont {Gaburro}}]{Yu11Science}%
  \BibitemOpen
  \bibfield  {author} {\bibinfo {author} {\bibfnamefont {N.}~\bibnamefont {Yu}}, \bibinfo {author} {\bibfnamefont {P.}~\bibnamefont {Genevet}}, \bibinfo {author} {\bibfnamefont {M.~A.}\ \bibnamefont {Kats}}, \bibinfo {author} {\bibfnamefont {F.}~\bibnamefont {Aieta}}, \bibinfo {author} {\bibfnamefont {J.-P.}\ \bibnamefont {Tetienne}}, \bibinfo {author} {\bibfnamefont {F.}~\bibnamefont {Capasso}}, \ and\ \bibinfo {author} {\bibfnamefont {Z.}~\bibnamefont {Gaburro}},\ }\bibfield  {title} {\enquote {\bibinfo {title} {Light propagation with phase discontinuities: {G}eneralized laws of reflection and refraction},}\ }\href@noop {} {\bibfield  {journal} {\bibinfo  {journal} {Science}\ }\textbf {\bibinfo {volume} {334}},\ \bibinfo {pages} {333--337} (\bibinfo {year} {2011})}\BibitemShut {NoStop}%
\bibitem [{\citenamefont {Kamali}\ \emph {et~al.}(2018)\citenamefont {Kamali}, \citenamefont {Arbabi}, \citenamefont {Arbabi},\ and\ \citenamefont {Faraon}}]{Kamali18NanoP}%
  \BibitemOpen
  \bibfield  {author} {\bibinfo {author} {\bibfnamefont {S.~M.}\ \bibnamefont {Kamali}}, \bibinfo {author} {\bibfnamefont {E.}~\bibnamefont {Arbabi}}, \bibinfo {author} {\bibfnamefont {A.}~\bibnamefont {Arbabi}}, \ and\ \bibinfo {author} {\bibfnamefont {A.}~\bibnamefont {Faraon}},\ }\bibfield  {title} {\enquote {\bibinfo {title} {A review of dielectric optical metasurfaces for wavefront control},}\ }\href@noop {} {\bibfield  {journal} {\bibinfo  {journal} {Nanophotonics}\ }\textbf {\bibinfo {volume} {7}},\ \bibinfo {pages} {1041--1068} (\bibinfo {year} {2018})}\BibitemShut {NoStop}%
\bibitem [{\citenamefont {Arriz{\'o}n}\ \emph {et~al.}(2007)\citenamefont {Arriz{\'o}n}, \citenamefont {Ruiz}, \citenamefont {Carrada},\ and\ \citenamefont {Gonz{\'a}lez}}]{Arrizon07JOSAA}%
  \BibitemOpen
  \bibfield  {author} {\bibinfo {author} {\bibfnamefont {V.}~\bibnamefont {Arriz{\'o}n}}, \bibinfo {author} {\bibfnamefont {U.}~\bibnamefont {Ruiz}}, \bibinfo {author} {\bibfnamefont {R.}~\bibnamefont {Carrada}}, \ and\ \bibinfo {author} {\bibfnamefont {L.~A.}\ \bibnamefont {Gonz{\'a}lez}},\ }\bibfield  {title} {\enquote {\bibinfo {title} {Pixelated phase computer holograms for the accurate encoding of scalar complex fields},}\ }\href@noop {} {\bibfield  {journal} {\bibinfo  {journal} {J. Opt. Soc. Am. A}\ }\textbf {\bibinfo {volume} {24}},\ \bibinfo {pages} {3500--3507} (\bibinfo {year} {2007})}\BibitemShut {NoStop}%
\bibitem [{\citenamefont {Hall}\ and\ \citenamefont {Abouraddy}(2023)}]{Hall23NatPhys}%
  \BibitemOpen
  \bibfield  {author} {\bibinfo {author} {\bibfnamefont {L.~A.}\ \bibnamefont {Hall}}\ and\ \bibinfo {author} {\bibfnamefont {A.~F.}\ \bibnamefont {Abouraddy}},\ }\bibfield  {title} {\enquote {\bibinfo {title} {Observation of optical {de B}roglie-{M}ackinnon wave packets},}\ }\href@noop {} {\bibfield  {journal} {\bibinfo  {journal} {Nat. Phys.}\ }\textbf {\bibinfo {volume} {19}},\ \bibinfo {pages} {435--444} (\bibinfo {year} {2023})}\BibitemShut {NoStop}%
\bibitem [{\citenamefont {Kent}\ \emph {et~al.}(2021)\citenamefont {Kent}, \citenamefont {Reynolds}, \citenamefont {Raftrey}, \citenamefont {Campbell}, \citenamefont {Virasawmy}, \citenamefont {Dhuey}, \citenamefont {Chopdekar}, \citenamefont {Hierro-Rodriguez}, \citenamefont {Sorrentino}, \citenamefont {Pereiro}, \citenamefont {S.~Ferrer},\ and\ \citenamefont {Fischer}}]{Kent2021NC}%
  \BibitemOpen
  \bibfield  {author} {\bibinfo {author} {\bibfnamefont {N.}~\bibnamefont {Kent}}, \bibinfo {author} {\bibfnamefont {N.}~\bibnamefont {Reynolds}}, \bibinfo {author} {\bibfnamefont {D.}~\bibnamefont {Raftrey}}, \bibinfo {author} {\bibfnamefont {I.~T.~G.}\ \bibnamefont {Campbell}}, \bibinfo {author} {\bibfnamefont {S.}~\bibnamefont {Virasawmy}}, \bibinfo {author} {\bibfnamefont {S.}~\bibnamefont {Dhuey}}, \bibinfo {author} {\bibfnamefont {R.~V.}\ \bibnamefont {Chopdekar}}, \bibinfo {author} {\bibfnamefont {A.}~\bibnamefont {Hierro-Rodriguez}}, \bibinfo {author} {\bibfnamefont {A.}~\bibnamefont {Sorrentino}}, \bibinfo {author} {\bibfnamefont {E.}~\bibnamefont {Pereiro}}, \bibinfo {author} {\bibfnamefont {P.~Sutcliffe}\ \bibnamefont {S.~Ferrer}, \bibfnamefont {F.~Hellman}}, \ and\ \bibinfo {author} {\bibfnamefont {P.}~\bibnamefont {Fischer}},\ }\bibfield  {title} {\enquote {\bibinfo {title} {Creation and observation of hopfions in magnetic multilayer systems},}\ }\href@noop {} {\bibfield  {journal} {\bibinfo
  {journal} {Nat. Comm.}\ }\textbf {\bibinfo {volume} {12}},\ \bibinfo {pages} {1562} (\bibinfo {year} {2021})}\BibitemShut {NoStop}%
\bibitem [{\citenamefont {Bliokh}\ \emph {et~al.}(2015{\natexlab{a}})\citenamefont {Bliokh}, \citenamefont {Rodr{\'\i}guez-Fortu{\~n}o}, \citenamefont {Nori},\ and\ \citenamefont {Zayats}}]{Bliokh2015NP}%
  \BibitemOpen
  \bibfield  {author} {\bibinfo {author} {\bibfnamefont {K.~Yu.}\ \bibnamefont {Bliokh}}, \bibinfo {author} {\bibfnamefont {F.~J.}\ \bibnamefont {Rodr{\'\i}guez-Fortu{\~n}o}}, \bibinfo {author} {\bibfnamefont {F.}~\bibnamefont {Nori}}, \ and\ \bibinfo {author} {\bibfnamefont {A.~V.}\ \bibnamefont {Zayats}},\ }\bibfield  {title} {\enquote {\bibinfo {title} {Spin-orbit interactions of light},}\ }\href@noop {} {\bibfield  {journal} {\bibinfo  {journal} {Nat. Phot.}\ }\textbf {\bibinfo {volume} {9}},\ \bibinfo {pages} {796--808} (\bibinfo {year} {2015}{\natexlab{a}})}\BibitemShut {NoStop}%
\bibitem [{\citenamefont {Bliokh}\ \emph {et~al.}(2015{\natexlab{b}})\citenamefont {Bliokh}, \citenamefont {Smirnova},\ and\ \citenamefont {Nori}}]{Bliokh2015Science}%
  \BibitemOpen
  \bibfield  {author} {\bibinfo {author} {\bibfnamefont {K.~Y.}\ \bibnamefont {Bliokh}}, \bibinfo {author} {\bibfnamefont {D.}~\bibnamefont {Smirnova}}, \ and\ \bibinfo {author} {\bibfnamefont {F.}~\bibnamefont {Nori}},\ }\bibfield  {title} {\enquote {\bibinfo {title} {Quantum spin {H}all effect of light},}\ }\href@noop {} {\bibfield  {journal} {\bibinfo  {journal} {Science}\ }\textbf {\bibinfo {volume} {348}},\ \bibinfo {pages} {1448--1451} (\bibinfo {year} {2015}{\natexlab{b}})}\BibitemShut {NoStop}%
\bibitem [{\citenamefont {Rego}\ \emph {et~al.}(2019)\citenamefont {Rego}, \citenamefont {Dorney}, \citenamefont {Brooks}, \citenamefont {Nguyen}, \citenamefont {Liao}, \citenamefont {{San R}om{\'a}n}, \citenamefont {Couch}, \citenamefont {Liu}, \citenamefont {Pisanty}, \citenamefont {Lewenstein}, \citenamefont {Plaja1}, \citenamefont {Kapteyn}, \citenamefont {Murnane},\ and\ \citenamefont {Hern{\'a}ndez-Garc{\'i}a}}]{Rego2019Science}%
  \BibitemOpen
  \bibfield  {author} {\bibinfo {author} {\bibfnamefont {L.}~\bibnamefont {Rego}}, \bibinfo {author} {\bibfnamefont {K.~M.}\ \bibnamefont {Dorney}}, \bibinfo {author} {\bibfnamefont {N.~J.}\ \bibnamefont {Brooks}}, \bibinfo {author} {\bibfnamefont {Q.~L.}\ \bibnamefont {Nguyen}}, \bibinfo {author} {\bibfnamefont {C.}~\bibnamefont {Liao}}, \bibinfo {author} {\bibfnamefont {J.}~\bibnamefont {{San R}om{\'a}n}}, \bibinfo {author} {\bibfnamefont {D.~E.}\ \bibnamefont {Couch}}, \bibinfo {author} {\bibfnamefont {A.}~\bibnamefont {Liu}}, \bibinfo {author} {\bibfnamefont {E.}~\bibnamefont {Pisanty}}, \bibinfo {author} {\bibfnamefont {M.}~\bibnamefont {Lewenstein}}, \bibinfo {author} {\bibfnamefont {L.}~\bibnamefont {Plaja1}}, \bibinfo {author} {\bibfnamefont {H.~C.}\ \bibnamefont {Kapteyn}}, \bibinfo {author} {\bibfnamefont {M.~M.}\ \bibnamefont {Murnane}}, \ and\ \bibinfo {author} {\bibfnamefont {C.}~\bibnamefont {Hern{\'a}ndez-Garc{\'i}a}},\ }\bibfield  {title} {\enquote {\bibinfo {title} {Generation of
  extreme-ultraviolet beams with time-varying orbital angular momentum},}\ }\href@noop {} {\bibfield  {journal} {\bibinfo  {journal} {Science}\ }\textbf {\bibinfo {volume} {364}},\ \bibinfo {pages} {eaaw9486} (\bibinfo {year} {2019})}\BibitemShut {NoStop}%
\bibitem [{\citenamefont {Chen}\ \emph {et~al.}(2022)\citenamefont {Chen}, \citenamefont {Zhu}, \citenamefont {Huo}, \citenamefont {Song}, \citenamefont {Lezec}, \citenamefont {Xu},\ and\ \citenamefont {Agrawal}}]{Chen2022AdvSc}%
  \BibitemOpen
  \bibfield  {author} {\bibinfo {author} {\bibfnamefont {L.}~\bibnamefont {Chen}}, \bibinfo {author} {\bibfnamefont {W.}~\bibnamefont {Zhu}}, \bibinfo {author} {\bibfnamefont {P.}~\bibnamefont {Huo}}, \bibinfo {author} {\bibfnamefont {J.}~\bibnamefont {Song}}, \bibinfo {author} {\bibfnamefont {H.~J.}\ \bibnamefont {Lezec}}, \bibinfo {author} {\bibfnamefont {T.}~\bibnamefont {Xu}}, \ and\ \bibinfo {author} {\bibfnamefont {A.}~\bibnamefont {Agrawal}},\ }\bibfield  {title} {\enquote {\bibinfo {title} {Synthesizing ultrafast optical pulses with arbitrary spatiotemporal control},}\ }\href@noop {} {\bibfield  {journal} {\bibinfo  {journal} {Sci. Adv.}\ }\textbf {\bibinfo {volume} {8}},\ \bibinfo {pages} {eabq8314} (\bibinfo {year} {2022})}\BibitemShut {NoStop}%
\bibitem [{\citenamefont {Su}\ \emph {et~al.}(2023)\citenamefont {Su}, \citenamefont {Zou}, \citenamefont {Zhou}, \citenamefont {Song}, \citenamefont {Wang}, \citenamefont {Zeng}, \citenamefont {Jiang}, \citenamefont {Duan}, \citenamefont {Karpov}, \citenamefont {Kippenberg}, \citenamefont {Tur}, \citenamefont {Christodoulides},\ and\ \citenamefont {Willner}}]{Su23ArXiV}%
  \BibitemOpen
  \bibfield  {author} {\bibinfo {author} {\bibfnamefont {X.}~\bibnamefont {Su}}, \bibinfo {author} {\bibfnamefont {K.}~\bibnamefont {Zou}}, \bibinfo {author} {\bibfnamefont {H.}~\bibnamefont {Zhou}}, \bibinfo {author} {\bibfnamefont {H.}~\bibnamefont {Song}}, \bibinfo {author} {\bibfnamefont {Y.}~\bibnamefont {Wang}}, \bibinfo {author} {\bibfnamefont {R.}~\bibnamefont {Zeng}}, \bibinfo {author} {\bibfnamefont {Z.}~\bibnamefont {Jiang}}, \bibinfo {author} {\bibfnamefont {Y.}~\bibnamefont {Duan}}, \bibinfo {author} {\bibfnamefont {M.}~\bibnamefont {Karpov}}, \bibinfo {author} {\bibfnamefont {T.~J.}\ \bibnamefont {Kippenberg}}, \bibinfo {author} {\bibfnamefont {M.}~\bibnamefont {Tur}}, \bibinfo {author} {\bibfnamefont {D.~N.}\ \bibnamefont {Christodoulides}}, \ and\ \bibinfo {author} {\bibfnamefont {A.~E.}\ \bibnamefont {Willner}},\ }\bibfield  {title} {\enquote {\bibinfo {title} {Temporally and longitudinally tailored dynamic space-time wave packets},}\ }\href@noop {} {\bibfield  {journal} {\bibinfo  {journal}
  {arXiv:2308.11213}\ } (\bibinfo {year} {2023})}\BibitemShut {NoStop}%
\bibitem [{\citenamefont {Dai}\ \emph {et~al.}(2022)\citenamefont {Dai}, \citenamefont {Ghosh}, \citenamefont {Yang}, \citenamefont {Zhou}, \citenamefont {Huang},\ and\ \citenamefont {Petek}}]{Dai2022NatRevPhys}%
  \BibitemOpen
  \bibfield  {author} {\bibinfo {author} {\bibfnamefont {Y.}~\bibnamefont {Dai}}, \bibinfo {author} {\bibfnamefont {A.}~\bibnamefont {Ghosh}}, \bibinfo {author} {\bibfnamefont {S.}~\bibnamefont {Yang}}, \bibinfo {author} {\bibfnamefont {Z.}~\bibnamefont {Zhou}}, \bibinfo {author} {\bibfnamefont {C.}~\bibnamefont {Huang}}, \ and\ \bibinfo {author} {\bibfnamefont {H.}~\bibnamefont {Petek}},\ }\bibfield  {title} {\enquote {\bibinfo {title} {Poincar{\'e} engineering of surface plasmon polaritons},}\ }\href@noop {} {\bibfield  {journal} {\bibinfo  {journal} {Nat. Rev. Phys.}\ }\textbf {\bibinfo {volume} {4}},\ \bibinfo {pages} {562--564} (\bibinfo {year} {2022})}\BibitemShut {NoStop}%
\bibitem [{\citenamefont {Berruto}\ \emph {et~al.}(2018)\citenamefont {Berruto}, \citenamefont {Madan}, \citenamefont {Murooka}, \citenamefont {Vanacore}, \citenamefont {Pomarico}, \citenamefont {Rajeswari}, \citenamefont {Lamb}, \citenamefont {Huang}, \citenamefont {Kruchkov}, \citenamefont {Togawa}, \citenamefont {LaGrange}, \citenamefont {McGrouther}, \citenamefont {R\o{}nnow},\ and\ \citenamefont {Carbone}}]{Beruto2018PRL}%
  \BibitemOpen
  \bibfield  {author} {\bibinfo {author} {\bibfnamefont {G.}~\bibnamefont {Berruto}}, \bibinfo {author} {\bibfnamefont {I.}~\bibnamefont {Madan}}, \bibinfo {author} {\bibfnamefont {Y.}~\bibnamefont {Murooka}}, \bibinfo {author} {\bibfnamefont {G.~M.}\ \bibnamefont {Vanacore}}, \bibinfo {author} {\bibfnamefont {E.}~\bibnamefont {Pomarico}}, \bibinfo {author} {\bibfnamefont {J.}~\bibnamefont {Rajeswari}}, \bibinfo {author} {\bibfnamefont {R.}~\bibnamefont {Lamb}}, \bibinfo {author} {\bibfnamefont {P.}~\bibnamefont {Huang}}, \bibinfo {author} {\bibfnamefont {A.~J.}\ \bibnamefont {Kruchkov}}, \bibinfo {author} {\bibfnamefont {Y.}~\bibnamefont {Togawa}}, \bibinfo {author} {\bibfnamefont {T.}~\bibnamefont {LaGrange}}, \bibinfo {author} {\bibfnamefont {D.}~\bibnamefont {McGrouther}}, \bibinfo {author} {\bibfnamefont {H.~M.}\ \bibnamefont {R\o{}nnow}}, \ and\ \bibinfo {author} {\bibfnamefont {F.}~\bibnamefont {Carbone}},\ }\bibfield  {title} {\enquote {\bibinfo {title} {Laser-induced skyrmion writing and erasing in
  an ultrafast cryo-{L}orentz transmission electron microscope},}\ }\href@noop {} {\bibfield  {journal} {\bibinfo  {journal} {Phys. Rev. Lett.}\ }\textbf {\bibinfo {volume} {120}},\ \bibinfo {pages} {117201} (\bibinfo {year} {2018})}\BibitemShut {NoStop}%
\bibitem [{\citenamefont {Basini}\ \emph {et~al.}(2024)\citenamefont {Basini}, \citenamefont {Pancaldi}, \citenamefont {Wehinger}, \citenamefont {Udina}, \citenamefont {Unikandanunni}, \citenamefont {Tadano}, \citenamefont {Hoffmann}, \citenamefont {Balatsky},\ and\ \citenamefont {Bonetti}}]{Basini2024Nat}%
  \BibitemOpen
  \bibfield  {author} {\bibinfo {author} {\bibfnamefont {M.}~\bibnamefont {Basini}}, \bibinfo {author} {\bibfnamefont {M.}~\bibnamefont {Pancaldi}}, \bibinfo {author} {\bibfnamefont {B.}~\bibnamefont {Wehinger}}, \bibinfo {author} {\bibfnamefont {M.}~\bibnamefont {Udina}}, \bibinfo {author} {\bibfnamefont {V.}~\bibnamefont {Unikandanunni}}, \bibinfo {author} {\bibfnamefont {T.}~\bibnamefont {Tadano}}, \bibinfo {author} {\bibfnamefont {M.~C.}\ \bibnamefont {Hoffmann}}, \bibinfo {author} {\bibfnamefont {A.~V.}\ \bibnamefont {Balatsky}}, \ and\ \bibinfo {author} {\bibfnamefont {S.}~\bibnamefont {Bonetti}},\ }\bibfield  {title} {\enquote {\bibinfo {title} {Terahertz electric-field-driven dynamical multiferroicity in {SrTiO$_{3}$}},}\ }\href@noop {} {\bibfield  {journal} {\bibinfo  {journal} {Nature}\ }\textbf {\bibinfo {volume} {628}},\ \bibinfo {pages} {534–539} (\bibinfo {year} {2024})}\BibitemShut {NoStop}%
\bibitem [{\citenamefont {Ichiji}\ \emph {et~al.}(2023)\citenamefont {Ichiji}, \citenamefont {Oue}, \citenamefont {Yessenov}, \citenamefont {Schepler}, \citenamefont {Abouraddy},\ and\ \citenamefont {Kubo}}]{Ichiji2023PRA}%
  \BibitemOpen
  \bibfield  {author} {\bibinfo {author} {\bibfnamefont {N.}~\bibnamefont {Ichiji}}, \bibinfo {author} {\bibfnamefont {D.}~\bibnamefont {Oue}}, \bibinfo {author} {\bibfnamefont {M.}~\bibnamefont {Yessenov}}, \bibinfo {author} {\bibfnamefont {K.~L.}\ \bibnamefont {Schepler}}, \bibinfo {author} {\bibfnamefont {A.~F.}\ \bibnamefont {Abouraddy}}, \ and\ \bibinfo {author} {\bibfnamefont {A.}~\bibnamefont {Kubo}},\ }\bibfield  {title} {\enquote {\bibinfo {title} {Transverse spin angular momentum of a space-time surface plasmon polariton wave packet},}\ }\href@noop {} {\bibfield  {journal} {\bibinfo  {journal} {Phys. Rev. A}\ }\textbf {\bibinfo {volume} {107}},\ \bibinfo {pages} {063517} (\bibinfo {year} {2023})}\BibitemShut {NoStop}%
\bibitem [{\citenamefont {Wu}\ \emph {et~al.}(2022)\citenamefont {Wu}, \citenamefont {Yu}, \citenamefont {Z.Zhu}, \citenamefont {Gao}, \citenamefont {Ding}, \citenamefont {Zhou}, \citenamefont {Hu}, \citenamefont {Rosales-Guzm\'{a}n}, \citenamefont {Shen},\ and\ \citenamefont {Shi}}]{Wu:22}%
  \BibitemOpen
  \bibfield  {author} {\bibinfo {author} {\bibfnamefont {H.}~\bibnamefont {Wu}}, \bibinfo {author} {\bibfnamefont {B.}~\bibnamefont {Yu}}, \bibinfo {author} {\bibnamefont {Z.Zhu}}, \bibinfo {author} {\bibfnamefont {W.}~\bibnamefont {Gao}}, \bibinfo {author} {\bibfnamefont {D.}~\bibnamefont {Ding}}, \bibinfo {author} {\bibfnamefont {Z.}~\bibnamefont {Zhou}}, \bibinfo {author} {\bibfnamefont {X.}~\bibnamefont {Hu}}, \bibinfo {author} {\bibfnamefont {C.}~\bibnamefont {Rosales-Guzm\'{a}n}}, \bibinfo {author} {\bibfnamefont {Y.}~\bibnamefont {Shen}}, \ and\ \bibinfo {author} {\bibfnamefont {B.}~\bibnamefont {Shi}},\ }\bibfield  {title} {\enquote {\bibinfo {title} {Conformal frequency conversion for arbitrary vectorial structured light},}\ }\href@noop {} {\bibfield  {journal} {\bibinfo  {journal} {Optica}\ }\textbf {\bibinfo {volume} {9}},\ \bibinfo {pages} {187--196} (\bibinfo {year} {2022})}\BibitemShut {NoStop}%
\bibitem [{\citenamefont {Shiri}\ \emph {et~al.}(2020)\citenamefont {Shiri}, \citenamefont {Yessenov}, \citenamefont {Webster}, \citenamefont {Schepler},\ and\ \citenamefont {Abouraddy}}]{Shiri20NC}%
  \BibitemOpen
  \bibfield  {author} {\bibinfo {author} {\bibfnamefont {A.}~\bibnamefont {Shiri}}, \bibinfo {author} {\bibfnamefont {M.}~\bibnamefont {Yessenov}}, \bibinfo {author} {\bibfnamefont {S.}~\bibnamefont {Webster}}, \bibinfo {author} {\bibfnamefont {K.~L.}\ \bibnamefont {Schepler}}, \ and\ \bibinfo {author} {\bibfnamefont {A.~F.}\ \bibnamefont {Abouraddy}},\ }\bibfield  {title} {\enquote {\bibinfo {title} {Hybrid guided space-time optical modes in unpatterned films},}\ }\href@noop {} {\bibfield  {journal} {\bibinfo  {journal} {Nat. Commun.}\ }\textbf {\bibinfo {volume} {11}},\ \bibinfo {pages} {6273} (\bibinfo {year} {2020})}\BibitemShut {NoStop}%
\bibitem [{\citenamefont {Diouf}\ \emph {et~al.}(2022)\citenamefont {Diouf}, \citenamefont {Lin}, \citenamefont {Harling},\ and\ \citenamefont {Toussaint}}]{Diouf22SA}%
  \BibitemOpen
  \bibfield  {author} {\bibinfo {author} {\bibfnamefont {M.}~\bibnamefont {Diouf}}, \bibinfo {author} {\bibfnamefont {Z.}~\bibnamefont {Lin}}, \bibinfo {author} {\bibfnamefont {M.}~\bibnamefont {Harling}}, \ and\ \bibinfo {author} {\bibfnamefont {K.~C.}\ \bibnamefont {Toussaint}},\ }\bibfield  {title} {\enquote {\bibinfo {title} {Demonstration of speckle resistance using space–time light sheets},}\ }\href@noop {} {\bibfield  {journal} {\bibinfo  {journal} {Sci. Rep.}\ }\textbf {\bibinfo {volume} {12}},\ \bibinfo {pages} {14064} (\bibinfo {year} {2022})}\BibitemShut {NoStop}%
\end{thebibliography}%

\clearpage

\clearpage

\clearpage

\clearpage

\end{document}